\documentclass[11pt,preprint]{aastex}

\received{}
\shortauthors {Shaw et al.}
\shorttitle {HST Images of Magellanic Cloud Planetary Nebulae} 

\begin{document}

\title {Hubble Space Telescope Images of Magellanic Cloud Planetary Nebulae\footnote
{Based on observations with the NASA/ESA Hubble Space Telescope, 
obtained at the Space Telescope Science Institute, which is operated 
by the Association of Universities for Research in Astronomy, Inc., under
NASA contract NAS 5-26555.} 
}

\author {Richard A.~Shaw, Letizia Stanghellini}
\affil {National Optical Astronomy Observatory, Tucson, AZ 85719}
\email {shaw@noao.edu, letizia@noao.edu}
\and
\author {Eva Villaver\altaffilmark{2}, Max Mutchler}
\affil {Space Telescope Science Institute, Baltimore, MD 21218}
\email {villaver@stsci.edu, mutchler@stsci.edu}

\altaffiltext{2}{Affiliated with the Hubble Space Telescope Space 
Department of ESA.}


\begin{abstract} 

We present images and slitless spectra which were obtained in 
{\it HST} surveys of Planetary Nebulae (PNe) in both the Large and Small 
Magellanic Clouds, using the Space Telescope Imaging Spectrograph.  
These new data on 59 PNe (54 in the LMC and five in the SMC) permit us to 
determine the nebular dimensions and morphology in the monochromatic 
light of several emission lines: H$\alpha$, [\ion{N}{2}] $\lambda$6583 and 
[\ion{O}{3}] $\lambda$5007, plus others of varying ionization, including 
[\ion{O}{1}], \ion{He}{1}, and [\ion{S}{2}]. We describe the nebular 
morphology and related features in detail. This survey, when combined 
with similar data from our prior {\it HST} programs and other archived 
PN images, brings the total of nebulae imaged with {\it HST} to 114 in 
the LMC and 35 in the SMC. We describe various basic properties for the 
sample, including sizes, morphologies, densities, and completeness. Trends in 
[\ion{O}{3}] $\lambda$5007 flux, surface brightness, and electron density 
with physical radius suggest that many nebulae, particularly those with 
bipolar morphology, may be optically thick even at large size. Bipolars 
also show the most extreme values of [\ion{N}{2}]/H$\alpha$ flux ratios, 
which is a rough indicator N enrichment. 

\end{abstract}


\section {Introduction} 

Planetary nebulae (PNe) in the Magellanic Clouds have been employed very 
successfully over the past few decades to investigate a number of very 
important astrophysical questions, including chemical abundances in the 
ISM of these galaxies, abundance yields of low-mass stars, the 
kinematical properties of the host galaxies, and even the extra-Galactic 
distance scale. In the past two decades a number of investigators have 
turned their attention to understanding the individual nebulae in detail 
in an effort to improve our understanding of stellar evolution, chemical 
enrichments, and the formation and evolution of the nebulae themselves. 
The effects of stellar evolution through the Asymptotic Giant Branch (AGB) 
on the Galactic PN abundances and on the central star (CS) mass and 
evolutionary paths have been well studied. However, recent modelling 
of the co-evolution of PNe and their central stars \citep{Vill_etal02, 
Perinotto_etal04} shows that the PN detectability and many of the 
observed properties of the PNe are intimately tied to the post-AGB 
evolution of the CS. 

Planetary nebulae are in many ways complex phenomena. Theories of their 
origin, morphology, chemical composition, evolution, and their use in 
inferring the evolutionary state of their CSs are complex and generally 
rich in their predictive power. Exploring their validity requires samples 
that are well populated, volume- and flux-complete, and uniformly observed 
in order to assure that the full set of observable phenomena are represented. 
For these purposes the study of Magellanic Cloud planetary nebulae (MCPNe) 
is potentially ideal: the Magellanic Clouds are massive enough that 
large numbers of PNe are available for study; the nebulae within a given 
Cloud all lie at essentially the same distance; the nebulae are bright 
enough that high quality spectra can be obtained easily, yet they are 
distant enough that the whole nebula can be fit within typical spectroscopic 
apertures; they are near enough that they can be resolved in considerable 
detail with the best resolution telescopes (such as {\it HST} or modern 
ground-based telescopes with adaptive optics); and the foreground 
extinction is very low. What has most limited the observational study of 
critical nebular and CS properties (and their correlation with nebular 
morphology) in the Magellanic Clouds up until the begining of this 
decade has been the comparatively small number of nebulae with high 
resolution images. 

{\it HST} imagery of MCPNe began 
with GTO program 1046 by \citet{Blades_etal92}, and was followed up by 
more than a dozen others, which are listed in Table~\ref{Surveys}. 
Reference keys to papers that analyze PN properties based on the {\it HST} 
programs are given in the last column of the Table. 
Unfortunately, the six earliest programs were executed 
with the spherically aberrated {\it HST}, prior to the first servicing 
mission, which limits the utility of the data.  
\citet{Stang_etal99} summarized the observations of the 17 LMC and 
10 SMC PNe from the first seven programs and expanded the analysis of 
these data to study the PN morphologies.  More comprehensive imaging surveys 
followed in the next few {\it HST} observing cycles, most of which were 
conducted as SNAPSHOT programs, as described by \citet[hereafter 
Paper~I]{Shaw_etal2001}, \citet[hereafter Paper~II]{Stang_etal02}, and 
\citet[hereafter Paper~III]{Stang_etal03}. These surveys expanded the 
available sample to 59 PNe in the LMC and 37 in the SMC, including some 
PNe from prior programs that were re-observed with STIS. 
The {\it HST} programs described in this paper, including one of LMC PNe 
(Cycle 9, Program 9077), and one of the SMC PNe (Cycle 13, Program 10251), 
extend earlier work and provide imagery on 54 LMC PNe and 
five SMC PNe. Other, mostly serendipitous {\it HST} observations from 
various programs have added images of another five LMC PNe and one 
additional SMC PN.  In sum, these contributions to the {\it HST} archive 
provide the most complete dataset on extra-galactic PN morphology 
achieved to date, with imaging of 114 unique PNe in the LMC and 35 
unique PNe in the SMC. 
The observational data, including the images and the emission line fluxes, 
are described in \S~2, and the nebular dimensions, morphology, and 
detailed features are described in \S~3. We analyze the LMC and SMC 
nebular properties in \S~4, and summarize our conclusions in \S~5. 

\section {Observational Data}

\subsection {Imaging} 

The data presented here from GO programs 9077 and 10251 were obtained with 
{\it HST} using the Space Telescope Imaging Spectrograph (STIS).  See 
\citet{Woodgate_etal98} for a description of the instrument, and 
\citet{Kimble_etal98} for a description of the initial, on-orbit performance.  
The observing approach and instrument configuration are almost identical 
to that described in Paper~I. Briefly, the targets were scheduled as 
SNAPSHOT exposures, which meant that the actual observations were selected 
on the basis of expediency for the telescope scheduling system, and on the 
availability of visits with suitable durations. 
All known, yet unobserved LMC PNe were potential targets for program 9077 
(i.e., 224 visits were granted for program 9077), though only a modest 
fraction were actually observed. 
In practice, brighter targets with short exposure times were 
observed much more often than fainter targets requiring longer visits. 
For program 10251 where 53 visits were approved (in order to observe 
all SMC PNe within $\sim$5 mag of the brightest), only five relatively 
bright targets were observed prior to the failure of the STIS instrument. 
This introduces a significant bias when attempting to draw conclusions 
about the full sample of LMC PNe (see \S~4). 

All of our observations were made with the CCD detector, using a gain of 
1 $e^-$/ADU.  Most of our exposures were split into two components of equal 
duration to facilitate cosmic ray removal.  We obtained slitless spectra 
with the G430M and G750M gratings which yielded monochromatic images of the 
target nebulae in several important emission lines.  
We also obtained direct, broad-band images with the clear (50CCD) aperture 
in order to measure central star (CS) magnitudes as faint as $V\approx25$ 
with a short exposure, and to distinguish between the spatial and the 
velocity structure in rapidly expanding nebulae. The fraction of CS 
detections with this technique is close to 65\% \citep{Villa03,Villa04,
Villa06}. The observing log is presented in Table~\ref{ObsLog}.  

The calibration procedure is nearly identical to that described in Paper~I, 
except that more recent versions of the processing software were 
employed. The instrumental signature was removed from the images and 
spectral images with the STIS calibration pipeline, version 2.18 
(2005 Jan.), as initially described by \citet{Hodge_etal98a} and 
\citet{Hodge_etal98b}. This portion of the processing substracts the bias, 
corrects for hot pixels by scaling and subtracting dark exposures, and 
divides by a normalized flat-field image to correct for pixel-to-pixel 
variations in detector sensitivity. The pipeline also combines multiple 
exposures, using an anti-coincidence technique to remove cosmic ray trails 
and other transient artifacts. 
A noteworthy improvement to this version of the pipeline, compared to 
that used for Papers I, II, and III, is the correction for charge transfer 
(in)efficiency \citep{BoGo03}, the effects of which grow slowly in time 
owing to the degradation of the STIS CCD detectors in flight. 

A recent search of the {\it HST} archive has uncovered a few images in 
the vicinity of known MCPNe, most of which were observed as a part of 
pure parallel programs. We have analyzed five additional nebulae (one of 
which was imaged during an acquisition for a STIS spectroscopic program) 
and include the results in this paper. The observing log of these other 
programs is presented in Table~\ref{Serendip}. These images, with the 
exception of that for LMC--MA~17, were calibrated with the standard 
reduction pipelines for the instruments used to obtain them. These 
pipelines perform essentially the same corrections as that for our STIS 
images: bias correction, dark scaling and subtraction, and flat-fielding. 
In addition, the ACS pipeline performs a geometric rectification. For the 
present purpose we are concerned primarily with determining the coordinates, 
sizes, and morphologies of these nebulae, so that the accuracy of the 
reductions is not so important. However we did combine images taken with 
the same filter, when available, to eliminate cosmic rays and to improve 
the signal-to-noise ratio for presentation. 

\subsection {Spectroscopy}

Extraction of one-dimensional spectra from the STIS two-dimensional spectral 
images proceeded similarly to that described in Paper~II. For this 
purpose we used the X1D module of the STIS pipeline \citep{McGrath99}; 
the key parameters that define the extraction (the position and size of 
the extraction window) are given in Table~\ref{SpExtr}. A noteworthy 
improvement to this module since Papers~II and III were published was the 
implementation of a correction for the slow degradation of the detector 
sensitivity with time \citep{StysBoGo04}. Prior to extraction, the 
two-dimensional images were filtered to remove image artifacts, such as 
the effects of extreme charge traps (pixel values below $-25$ electrons) 
and incomplete corrections for hot pixels (isolated pixels with values of a 
few hundred electrons above that for adjacent pixels) in the immediate 
vicinity of emission features. These corrections to tens to perhaps a few 
hundred pixels have no effect on the flux calibration, but do have the effect 
of improving the realized signal-to-noise ratio in the extracted spectra for 
weak emission features. 
The fluxes from each of the detected emission lines were measured from the 
extracted spectra using the IRAF\footnote{IRAF 
is distributed by the National Optical Astronomy Observatory, which is 
operated by the Association of Universities for Research in Astronomy, 
Inc., under cooperative agreement with the National Science Foundation.}
task {\bf splot}. All of the fluxes were measured via direct integration 
above the local continuum, rather than assuming an intrinsic (e.g., 
Gaussian) shape to the emission profile, because of the often complex 
spatial profile of the nebulae. 

The emission line intensities are presented in Table~\ref{Flux}, uncorrected 
for interstellar reddening, and are normalized to F(H$\beta$)=100. Column~1 
gives the nebula name using the widely-used designation of \citet{SMP78} 
where possible, and otherwise by the original discovery catalog \citep{Jacoby80, 
Sa84, MG92, Mo94}. Column 2 gives the log of the flux in H$\beta$ 
$\lambda$4861 in erg cm$^{-2}$ s$^{-1}$, and column 3 gives the extinction 
constant, {\it c} (the logarithmic extinction at H$\beta$) as computed from our 
F(H$\alpha$)/F(H$\beta$) ratio and assuming an electron temperature of 
10,000~K and density of 10$^4$ cm$^{-3}$. The intensities for the bright nebular 
emission lines, relative to H$\beta = 100$, are given in columns 4 through 14. 
Blends of emission lines that occur for large nebulae are noted in the table, 
in which case the sum of the line intensities in the blend is given. 

\begin {figure}
\epsscale{0.6}
\includegraphics[angle=-90]{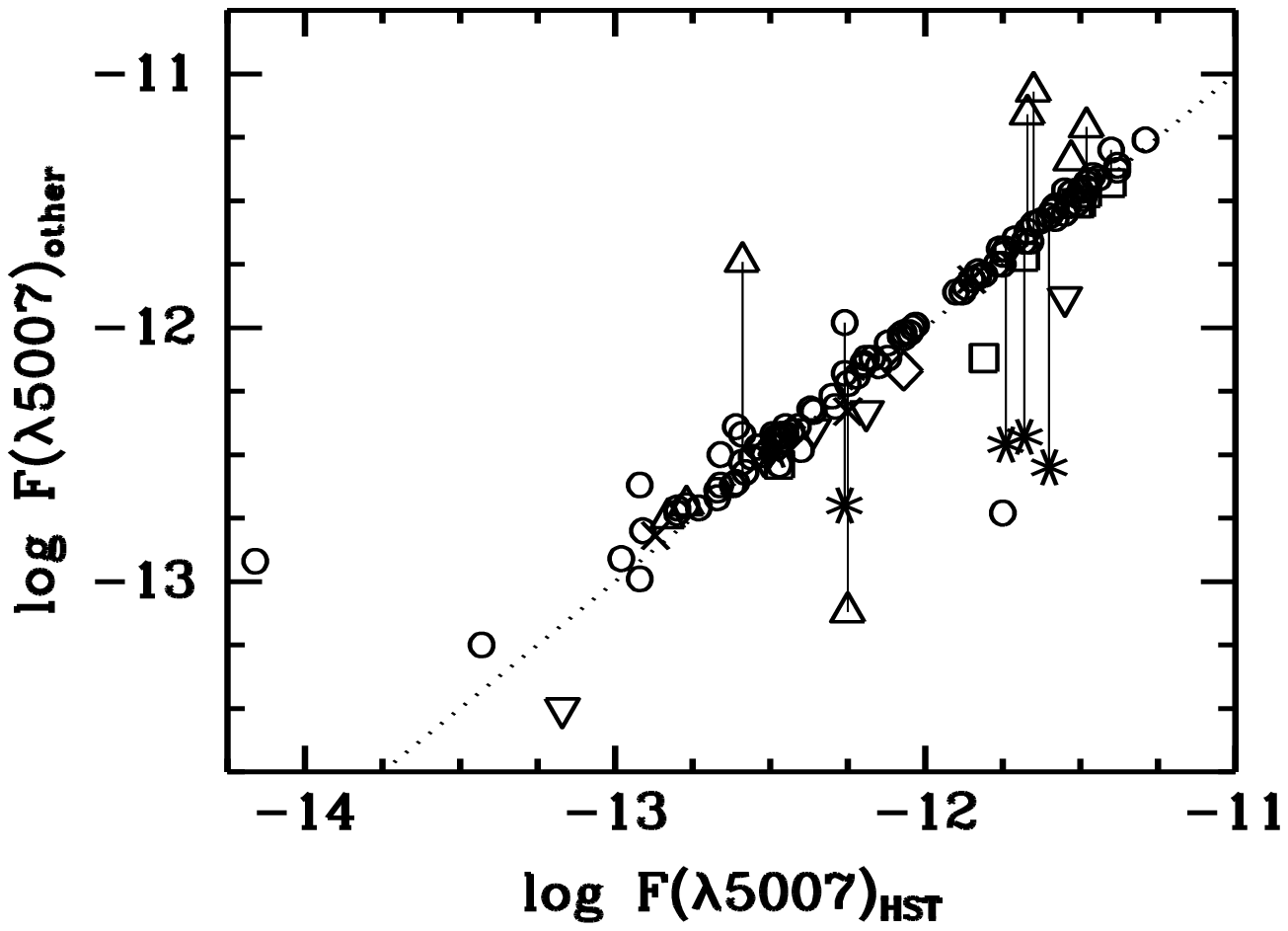}
\figcaption[f1.eps]{ 
Comparison of the log flux in [\ion{O}{3}] $\lambda$5007 (in 
erg cm$^{-2}$ s$^{-1}$) of various 
published values and those reported in this paper and in Papers~II and III, 
with the 1:1 relation indicated ({\it dotted line}). 
Symbol key: \citet{JWC90} ({\it circles}); \citet{JK93} ({\it squares}); 
\citet{BL89} ({\it triangles}); \citet{Vass_etal92} ({\it inverted triangles}); 
\citet{Wood_etal87} ({\it diamonds}); \citet{Meyss95} ({\it crosses}); 
\citet{Webster76} ({\it asterisks}). Examples of objects observed by more than 
one other investigator are connected with thin vertical lines. 
\label{FlxCmp}}
\end {figure}

We expect the accuracy and associated uncertainties of our relative fluxes 
to be very similar to those described in Papers~II and III, since the 
techniques for spectral extraction, calibration and flux measurement are 
nearly identical. Since the sample of Magellanic Cloud  PNe presented 
here and in Papers~II and III is now fairly large it is useful to compare 
our absolute fluxes to those in the literature, all of which are from 
ground-based data. We present in Figure~\ref{FlxCmp} a comparison of the 
published fluxes with ours in the [\ion{O}{3}] $\lambda$5007 emission line 
(scaling by the published absolute flux in H$\beta$ where necessary). The 
sample of MCPNe in common spans about two orders of magnitude in flux. 
Usually the agreement is very good, though in several cases (particularly 
for log fluxes less than $-12.75$) the agreement is poor. Some objects 
which were observed by more than two investigators are connected by thin 
vertical lines. In these cases it is clear that most of the most discrepant 
values are inaccurate. 

\begin {figure}
\epsscale{0.6}
\includegraphics[angle=-90]{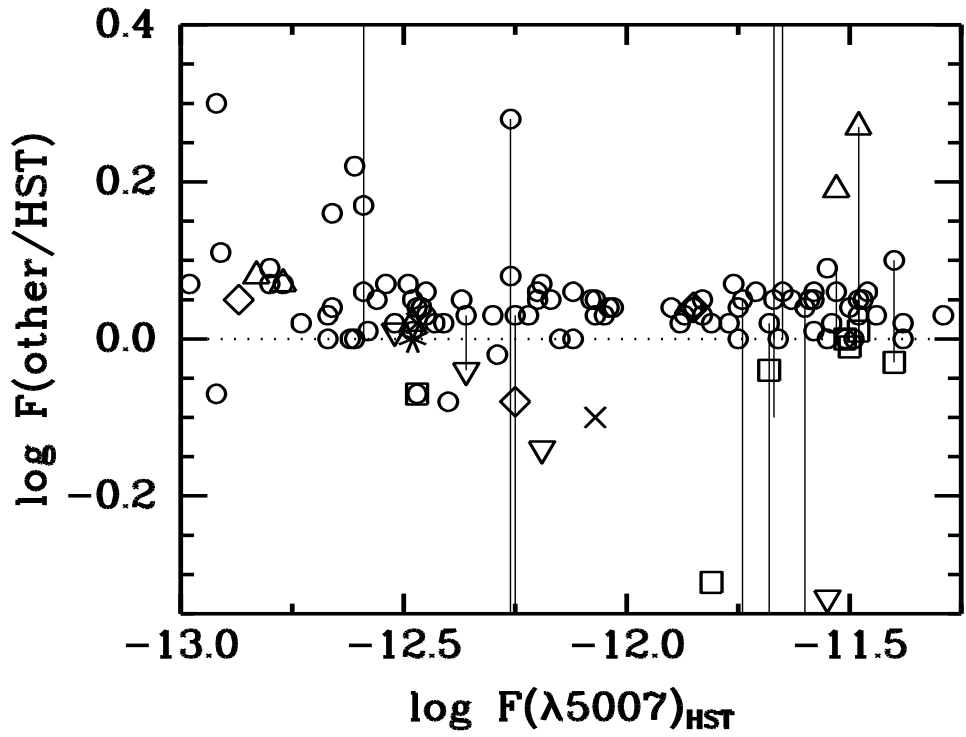}
\figcaption[f2.eps]{ 
Same as Fig.~\ref{FlxCmp}, except the ordinate is the log of the ratio of 
the flux in [\ion{O}{3}] between various published values and those 
reported in this paper and in Papers~II and III. Symbols are as in 
Fig~\ref{FlxCmp}. 
\label{FlxCmpB}}
\end {figure}

Of all the published fluxes with targets in common, those of \citet{BL89} 
cover the largest dynamic range. On the whole the agreement is good, but 
the deviations for individual nebulae are sometimes more than a factor of 
three, almost independent of the brightness. Objects in common with 
\citet{Meyss95} also show good agreement over a wide range, except for 
one extremely discrepant object (SMC--J~23) where the difference is a 
factor of 100. 
The greatest overlap of objects is with \citet[hereafter, JWC]{JWC90} 
where the agreement for objects brighter than $-13$ in the log is 
excellent, apart from a consistent offset of about 10\%, as shown in 
Figure~\ref{FlxCmpB}. The cause of the offset, where the JWC fluxes are 
consistently brighter than ours, is not entirely clear. For ground-based 
data the principal challenges to accuracy are the photometricity of the 
observing conditions, and relatively large entrance apertures that can 
admit a considerable amount of light from background sources. The light 
from the background includes field stars and (often) significant diffuse 
emission line radiation that varies with position. The STIS spectra are 
rather less susceptible to these particular effects, owing to the superior 
spatial resolution and the large 
effective entrance aperture (nearly 1\arcmin\ by 1\arcmin) which both 
assures that no nebular light is lost, while spreading the background 
diffuse emission over the whole detector. It is possible to exclude light 
inadvertantly from the STIS spectra by using virtual extraction slits 
that are too small. In this case we would expect to see fluxes that agree 
with those of JWC for small, nearly unresolved nebulae (owing to the well 
calibrated correction in the STIS pipeline for finite-aperature effects), 
but that become systematically too small for large nebulae, which we do 
not find. 
In any case, the systematic offset is far larger than uncertainty in the 
measurement errors claimed by JWC, and also larger than that for the 
absolute accuracy of STIS in this mode (a few percent). We conclude 
that the fluxes 
from this paper, Papers~II and III in this series, and those of 
\citet{JWC90} and \citet{JK93} are the most consistent and likely the most 
reliable for log fluxes greater than $-12.75$; below that value our fluxes 
are probably the most reliable to date. 

\section {Dimensions and Morphology}

The broad-band, [\ion{O}{3}] $\lambda$5007, and H$\alpha$ $\lambda$6563 + 
[\ion{N}{2}] $\lambda\lambda$6548, 6583 images for the target nebulae of 
program 9077 are presented in Figures~\ref{Neb_1} through \ref{Neb_13}, and 
in Figure~\ref{Neb_14} for the SMC nebulae of program 10251. Images of PNe 
from the other {\it HST} programs are presented in Figure~\ref{Neb_15}. 
All of the data are rendered as grey-scale images, usually with a log 
intensity scale in order to bring out the often faint structural 
features.\footnote{ 
Cut-out, false-color images of the entire {\it HST} sample are available 
for browsing at the MAST archive, at http://archive.stsci.edu/hst/mcpn/.} 
We classified the morphologies in our sample on the basis of the 
[\ion{O}{3}] $\lambda$5007 monochromatic images in the G430M spectra 
(although we were guided by the H$\alpha$ and [\ion{N}{2}] $\lambda$6583 
images), using the same classification scheme as in Papers I and III. 
The morphological types are: round (R), elliptical (E), quadrupolar (Q), 
point-symmetric (P), bipolar (B), and the sub-class of bipolar-core (BC). 
The BC designation is applied to 
nebulae that have a round or elliptical outer contour, but also have internal 
structure in the form of two lobes of emission that surround the center, and 
that are at least 20\% brighter than the immediately surrounding nebula. 
When present, such structure is an important morphological feature, in that 
these nebulae may be more closely related to pure bipolar (B) PNe than 
either R or possibly E (see \S~4, below). Our distinction between E and R 
was based (somewhat arbitrarily) on whether the major axis of the 
10\% intensity contour exceeded the minor axis by more than 10\%. 
Other important structural features, such as attached shells, ansae, and 
jets were noted as well. 
The morphological class was assigned primarily on the basis of structures 
evident in the [\ion{O}{3}] image, and would seldom have been different had 
we instead used another, lower-ionization line such as 
[\ion{N}{2}] $\lambda$6583. Indeed, in only one case (out of 61 resolved 
nebulae), for LMC--SMP~62, might the classification have been different. 
We conclude that at least for the broad morphological characteristics 
considered here, only the [\ion{O}{3}] $\lambda$5007 image is essential 
for a correct classification with this schema. A refined classification to quantify 
the effects of projection may now be possible \citep{ZhangKwok98}, but 
a significant improvement would require spatially resolved, high-resolution 
spectra to understand the velocity field of the gas. 

Table~\ref{Morph} gives detailed information for the PNe presented here. 
The sky coordinates are given in columns 2 and 3, and the photometric 
radii are given in column 4, which were derived according to the method 
described by \citet{Stang_etal99}. The photometric radius, R$_{phot}$, 
corresponds to the size of a circular aperture that contains 85\% of the 
flux in [\ion{O}{3}] $\lambda$5007. R$_{phot}$ gives an objective 
measurement of nebular angular size which is insensitive to the S/N 
ratio of the image, and is useful for evolution studies. 
The nebular diameters, given in column 5, were measured with respect to 
the 10\% intensity contour of the outermost structure, and are useful for 
conducting follow-on observations of the PNe.  The morphological 
classification is given in column 6. 

\subsection {Individual Nebulae}

As discussed in the previous section, we have observed 54 PNe in the LMC 
and five nebulae in the SMC. In addition, we have analyzed archived 
images of an additional five PNe in the LMC and one in the SMC. 
We describe in this subsection the morphological details and noteworthy 
spectral features for each nebula listed in Table~\ref{Morph}. We take 
special note of objects that show extreme ratios of 
F([\ion{N}{2}])/F(H$\alpha$): often this occurs for large objects. In these 
cases the lines are partially blended (see Table~\ref{Flux}), so that the 
ratio (or a limit) is determined only approximately from unblended portions 
of the emission. 

\begin {figure}
\epsscale{0.6}
\plotone {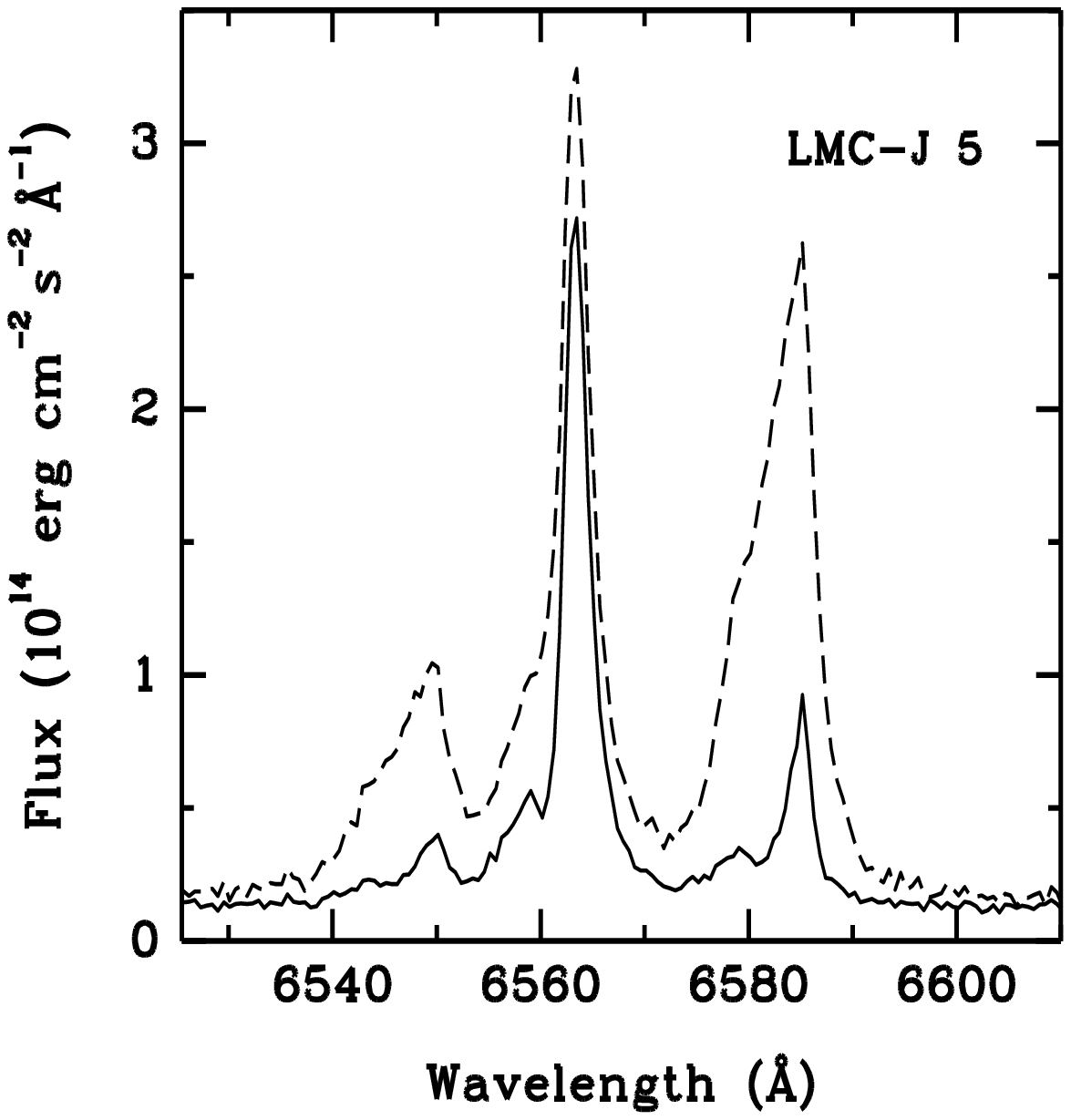}
\figcaption[f3.eps]{ 
Spectrum of LMC--J~5, comparing the emission of H$\alpha$ and [\ion{N}{2}] 
$\lambda\lambda$6548,6583 from the nebula plus star ({\it dashed}) to that 
of the star ({\it solid}) from a 3 pixel virtual extraction slit. 
\label{J5_spec}}
\end {figure}

{\it LMC--J~5.---}This nebula is an elliptical ring, though there is a knot 
of bright emission nearly coincident with the extremely bright central star 
(CS). The CS is displaced about 0\farcs25 (0.06 pc) along the major axis of 
the ellipse, near the eastern edge of the nebula, and might have been 
mistaken for a field star were it not for the associated emission. The core 
of the CS is saturated in the broad-band image, with a $V$ magnitude a 
little less than 17.6 \citep{Villa06}. The 
nebular emission from the [\ion{N}{2}] $\lambda$6583 line is a few times 
brighter than that from H$\alpha$, and a faint extension of nebulosity can 
be seen to the east. 
The surface brightness of the knot in the broad-band is roughly five 
times larger than that from the rest of the nebula. The emission spectrum 
of the knot is different from the rest of the nebula, and is mixed with 
stellar emission lines. Fig.~\ref{J5_spec} shows that the bulk H$\alpha$ 
emission is of stellar origin. A small deficit of emission blueward of the 
peak, which is also present in H$\beta$ but not in the forbidden lines, 
suggests a P-Cygni profile. \citet{Alves_etal97} found the CS to be 
irregularly variable, with an amplitude of up to 0.4 mag in the visible, 
and noted that the concomitant change in $V-R$ color was consistent 
with increased reddening. \citet{BL89} noted the presence of Fe emission 
lines in the spectrum. The combination of morphological, 
temporal, and spectral features suggests that J~5 is a genuine PN 
with a binary CS. 
(See Fig.~\ref{Neb_1}.)

{\it LMC--J~33.---}This slightly elliptical nebula has an attached outer 
shell that is almost twice the size and about one-third the surface 
brightness of the bright, inner nebula.  The central star is easily visible. 
(See Fig.~\ref{Neb_1}.)

{\it LMC--MA~17.---}This nebula has a quadrupolar morphology (which
is very rare, and is the same as that for SMP 27), with rough dimensions of
0.215 x 0.164 pc. It has what may be jets extending about a 0.3 pc from the 
central star, and are symmetric along an axis that is inclined roughly 30\degr 
from the minor axis. 
(See Fig.~\ref{Neb_16}.)

{\it LMC--MG~4.---}This is one of the largest nebulae in our sample, with 
very low average surface brightness that renders the morphological 
classification uncertain. The outer shape is elliptical, with a 
region of enhanced emission on the eastern edge that is a factor of three 
or more greater than the rest of the nebula. There is moderate emission in 
[\ion{O}{3}] and [\ion{S}{2}]; the emission near H$\alpha$ is dominated by 
[\ion{N}{2}], such that F($\lambda$6548+$\lambda$6583)/F($\lambda$6563) 
probably exceeds 5.  No central star is visible. 
(See Fig.~\ref{Neb_2}.)

{\it LMC--MG~14.---}This round nebula has a bipolar core surrounding a 
small central zone with reduced emission, and an attached outer shell that 
is twice as large as the primary shell, and about one-third as bright. 
The [\ion{O}{3}] emission is very strong, but the lower-ionization lines of 
[\ion{N}{2}], [\ion{S}{2}], [\ion{O}{1}], and \ion{He}{1} are absent. 
The CS is easily visible. 
(See Fig.~\ref{Neb_1}.)

{\it LMC--MG~16.---}This bipolar nebula shows a classic butterfly shape 
in the broad-band, with a pinched waist of roughly 0\farcs75, terminated 
by bright knots of emission. The emission in one``wing" extends at least 
1\farcs3 to the east of the very prominant CS. Much more structure is 
apparent in the red spectrum, where the strong [\ion{N}{2}] emission shows 
what appear to be a nearly orthogonal set of loops that intersect the 
bright emission knots at the ends of the waist. The emission in H$\alpha$ 
and the weak [\ion{O}{3}] lines is by comparison more uniform and compact, 
with no sign of loops. This is a case where the R$_{phot}$ as measured 
from [\ion{O}{3}] $\lambda$5007 would be significantly smaller had it 
been determined from [\ion{N}{2}] $\lambda$6583, and suggests stratified 
ionization. There is weak emission in [\ion{S}{2}] and possibly in 
[\ion{O}{1}]. 
(See Fig.~\ref{Neb_1}.)

{\it LMC--MG~29.---}This nebula is basically elliptical in shape, with 
additional structure. The broad-band image shows brightening on the ring 
edges to the SE and (to a lesser extent) the NW, and faint extensions 
to the NE and SW. A faint central star is also visible. The red spectrum 
reveals a striking degree of structure along the ring in the emission of 
[\ion{N}{2}] which is nearly absent in the brighter emission of H$\alpha$. 
The ring structure gives the appearance of braided strands. Emission in 
[\ion{O}{3}] is very strong, though [\ion{S}{2}] and [\ion{O}{1}] are 
present but weak. 
(See Fig.~\ref{Neb_1}.)

{\it LMC--MG~40.---}This is a very interesting elliptical nebula with a 
bipolar core and an attached outer shell. The pair of enhanced emission 
features within the core are roughly aligned with the major axis of the 
elliptical boundary, but are not located symmetrically along that axis. 
The core surface brightness is a factor of 3---4 brighter than that of 
the attached shell, which is also elliptical in shape but has a lower 
eccentricity (0.42 {\it vs.} 0.68). The limb of the outer shell is somewhat 
brightened on the eastern edge, which is suggestive of some kind of 
interaction with gas just outside of the illuminated shell. The CS is 
easily visible in the broad-band image. 
(See Fig.~\ref{Neb_3}.)

{\it LMC--MG~45.---}This nebula is barely resolved, but does show an 
elliptical shape in the emission lines. The nebular emission is very 
sharply peaked, and is difficult to distinguish from that of the CS. 
(See Fig.~\ref{Neb_3}.)

{\it LMC--MG~51.---}This nebula is slightly elliptical, with a bright core 
surrounded by a faint, attached shell. A central emission cavity is 
slightly offset from the center, but overall the core is a factor of 3---5 
brighter per unit area than the attached shell. The nebula is very bright in 
[\ion{O}{3}] and H$\alpha$, but [\ion{N}{2}] and the other low-ionization 
lines are not detected. The CS is very bright. 
(See Fig.~\ref{Neb_3}.)

{\it LMC--MG~60.---}This nebula is slightly elliptical with a fairly 
uniform surface brightness, apart from a small region of enhanced 
brightness to the NW of the bright CS. Beyond this brighter region the 
nebular emission drops off abruptly along a straight line, almost as if 
the nebular shell had run into an invisible wall. The image was taken 
through the ACS/WFC F775W filter, so the emission is probably dominated 
by H$\alpha$ and possibly [\ion{N}{2}]. There is a field star within the 
nebula, about 0\farcs5 from the CS and about 2 mag fainter. 
(See Fig.~\ref{Neb_16}.)

{\it LMC--MG~70.---}This nebula is elliptical in its outer contour, but the 
appearance in the light of H$\alpha$ and especially [\ion{N}{2}] is 
barrel-shaped, with a distinct bipolar core. There is no CS evident in 
the broad-band against the bright nebular emission. 
(See Fig.~\ref{Neb_3}.)

{\it LMC--Mo~7.---}This is a very strange, elliptical nebula with a very 
bright knot of emission on its southern edge. The emissions in H$\alpha$ 
and [\ion{O}{3}] show something of a ring feature, though the knot is 
relatively brighter in H$\alpha$. There is no trace of emission in 
[\ion{N}{2}] or any other low-ionization line. The CS is very bright in 
the broad-band. 
(See Fig.~\ref{Neb_3}.)

{\it LMC--Mo~17.---}This is an elliptical nebula with a compact core, 
surrounded by a much larger shell with a surface brightness that is two 
to three times fainter. The nebula is distinctly brighter in the ACS/WFC 
F814W frame than in F555W, which may be due to stronger emission in 
H$\alpha$ + [\ion{N}{2}] than in [\ion{O}{3}]. The CS might be detected 
but is very faint in both frames. There are a few comparatively bright 
field stars within the spatial extent of the nebula, including one within 
0\farcs9. 
(See Fig.~\ref{Neb_16}.)

{\it LMC--Mo~21.---}This bipolar nebula is so faint that only the ring-like 
waist is evident in the broad-band, along with what might be a faint CS. 
The emission in [\ion{N}{2}] is much brighter than H$\alpha$, such that 
F($\lambda$6548+$\lambda$6583)/F($\lambda$6563) probably exceeds 5.  The 
red spectral image shows more complicated structure in the waist, with 
faint emission extending perpendicular to the waist at least 2\farcs0. 
The nebula is barely detected in [\ion{O}{3}]. 
(See Fig.~\ref{Neb_2}.)

{\it LMC--Mo~33.---}This is an elliptical nebula with a central cavity, 
and relatively low average surface brightness. The outer boundary is rather 
sharply defined on the E and S limbs, and less so on the N and W. The 
H$\alpha$ emission is stronger than that from the overlapping [\ion{N}{2}]. 
The [\ion{O}{3}] emission suggests a somewhat bi-lobed ring structure. The 
CS is faint but apparent in the broad-band. 
(See Fig.~\ref{Neb_4}.)

{\it LMC--Mo~36.---}This is an elliptical nebula with fairly uniform 
surface brightness, apart from a fainter S limb, at the end of the major 
axis. The emission in [\ion{N}{2}] has a more sharply defined edge than 
that in H$\alpha$, and there is a suggestion of a ring or central cavity 
in the [\ion{N}{2}] lines. No CS is evident in the broad-band. 
(See Fig.~\ref{Neb_5}.)

{\it LMC--Mo~47.---}This nebula is extremely faint and the very low surface 
brightness makes it difficult to assess. It is probably round, and a very 
faint central star can be seen. It is detected in the emissions of 
H$\alpha$ and [\ion{O}{3}], but probably not in [\ion{N}{2}]. The derived 
fluxes are extremely uncertain, both because of the low intrinsic signal 
relative to the detector noise, and because the effects of charge transfer 
inefficiency are the most extreme for this case. 
(See Fig.~\ref{Neb_4}.)

{\it LMC--RP~265.---}The extremely low surface brightness of this object 
and the short exposure time with the WFPC2/F606W filter makes a 
morphological classification difficult, but this nebula is possibly a 
bipolar. No CS is evident in the image. 
(See Fig.~\ref{Neb_16}.)

{\it LMC--Sa~104a.---}This nebula is very bright, but is probably not 
resolved spatially. The emission lines are all extended in the dispersion 
direction (compared to spectra of other nebulae in this sample), however 
this is likely an artifact of marginally resolving the velocity profile 
of the gas expanding from the central source. In addition to the expected 
nebular lines, several other weak features are detected: some are 
unidentified, and some are \ion{He}{1} or \ion{He}{2} lines that could 
be nebular or stellar in origin. A deep, high-resolution spectrogram 
should be obtained to study the weak emission features. 
(See Fig.~\ref{Neb_5}.)

{\it LMC--Sa~107.---}This somewhat bizzare nebula is probably bipolar, 
though it might be point-symmetric. It 
has a variety of internal structures, including an incomplete ring of 
emission in the center of what might be a waist, with a variety of what 
look to be wisps or tails of emission extending at least 1\farcs7 away 
from the core, more or less symmetrically about the waist. The central 
cavity within the ring is broken at the N edge, and the bright CS is 
slightly displaced from the geometric center toward the break. The 
morphology is so disturbed that it is tempting to invoke jets or some other 
exotic shaping mechanism as an explanation. The emission in [\ion{N}{2}] 
is much brighter than H$\alpha$, and the [\ion{O}{3}] emission is weak. 
(See Fig.~\ref{Neb_5}.)

{\it LMC--Sa~117.---}This is a point-symmetric nebula, with knots of 
emission at each end of two separate axes that are not orthogonal, at least 
in projection. The knots are situated surround a cavity of emission that is 
slightly offset from the geometric center. The CS is easily visible. 
(See Fig.~\ref{Neb_5}.)

{\it LMC--Sa~121.---}This is a mildly elliptical nebula with a core of 
relatively constant surface brightness, surrounded by a shell with a 
surface brightness only a factor of two lower. Two ansae extend about 
0\farcs3 to the N and S from the outer shell. The morphology is more like 
a ring in the light of [\ion{N}{2}], which is no doubt due to stratification 
in the nebular ionization. The CS is detected but faint in the broad-band. 
(See Fig.~\ref{Neb_5}.)

{\it LMC--SMP~3.---}This nebula is barely resolved, but is probably round. 
The spatial extent of the nebular emission is strongly peaked, and is 
difficult to distinguish from the central star on the broad-band. 
Continuum emission is evident in the red spectrum, but may be entirely 
due to nebular continuum. 
(See Fig.~\ref{Neb_6}.)

{\it LMC--SMP~5.---}This is a mildly elliptical nebula with a single shell. 
The bright CS is easily detected in the broad-band, and can be seen 
also in the blue and red spectra. 
(See Fig.~\ref{Neb_6}.)

{\it LMC--SMP~6.---}This elliptical nebula has a slightly bipolar 
core. No CS is evident in the broad-band. 
(See Fig.~\ref{Neb_6}.)

{\it LMC--SMP~11.---}This nebula is one of the smallest known bipolar PNe in 
the LMC. The emission lobes are not as widely separated in the light of 
H$\alpha$ and [\ion{N}{2}] (less than 0\farcs2) as in [\ion{O}{3}] (roughly 
0\farcs35). Interestingly, there is a detached arc of emission located 2\farcs3, 
or 0.56 pc, from the center that extends roughly 40\degr\ from approximately 
S through SE. The arc is slightly more than 2\farcs0 long and 0\farcs4 wide, 
and has a surface brightness that is smaller than the main nebula by a 
factor of between 10 (in [\ion{O}{3}]) to 100 (in the broadband filter). (The arc 
lies outside the field of view of the figure, but it would not have been visible 
with the chosen intensity stretch.) 
The emission is apparent in the light of [\ion{O}{3}], and is just 
detected in the light of [\ion{N}{2}]. This arc is reminicent of the 
arc near SMP~27 described in Paper~I, and later shown by \citet{ReidParker06} 
to be part of a large, faint halo surrounding the core nebula. SMP~11 has 
a fairly high aggregate expansion velocity of 122 km/s \citep{Dopita_etal88}, 
with multiple velocity components. The CS is probably obscured by the bright 
nebular emission in the broad-band image. 
(See Fig.~\ref{Neb_6}.)

{\it LMC--SMP~14.---}This bipolar nebula shows a classic butterfly shape 
in the broad-band, with a pinched waist of roughly 1\farcs6, terminated 
by very bright knots of emission.  
The emission in [\ion{N}{2}] is much brighter than H$\alpha$, such that 
F($\lambda$6548+$\lambda$6583)/F($\lambda$6563) probably exceeds 5.  The 
[\ion{N}{2}] image shows more complicated structure in the wings, with 
faint emission extending perpendicular to the waist at least 1\farcs8, and 
surrounding a central cavity of diameter roughly the size of the waist. 
The nebula is fairly weak in [\ion{O}{3}]. 
At the center of the nebula is either a faint CS or a knot of emission. 
(See Fig.~\ref{Neb_7}.)

{\it LMC--SMP~29.---}This nebula is somewhat irregular, but with a 
bipolar core. It may be the smallest recognized bipolar to date 
in the LMC, with significantly enhanced emission in the E lobe. The 
bipolarity is even more pronounced in the [\ion{N}{2}] emission, where 
pairs of faint ansae can be seen extending to the S, N, and E by up to 
0\farcs7 from the center. No CS is evident within the bright nebulosity 
of the broad-band image. 
(See Fig.~\ref{Neb_6}.)

{\it LMC--SMP~37.---}This elliptical nebula shows enhanced emission on the 
N rim. The nearly complete ring of emission in the light of [\ion{N}{2}] 
suggests considerable stratification in the nebular ionization. 
No CS is detected. 
(See Fig.~\ref{Neb_8}.)

{\it LMC--SMP~39.---}This nebula bears a striking resemblence to SMP~37 
in structure, physical size, and its emission spectrum, although SMP~39 
emission shows a complete, unbroken ring. No CS is detected. 
(See Fig.~\ref{Neb_8}.)

{\it LMC--SMP~43.---}The morphology of this round, bipolar core nebula is 
strikingly similar to LMC--MG~40, with its elliptical core; faint, attached 
shell with a brighter limb (to the E), and a relatively prominant CS. As with 
MG~40, SMP~43 shows some intensity enhancement near each end of the major 
axis of the elliptically shaped core, especially in the light of [\ion{O}{3}]. 
(See Fig.~\ref{Neb_8}.)

{\it LMC--SMP~45.---}This nebula is probably bipolar, with an axis of 
symmetry running approximately N--S. Faint emission can be seen extending 
at least 2\farcs2 from the faint CS. The overall emission is 
highly irregular, though the [\ion{N}{2}] and [\ion{S}{2}] emission shows 
more wisps and a central region of reduced emission, or cavity, from 
low-ionization species. 
(See Fig.~\ref{Neb_7}.)

{\it LMC--SMP~47.---}This elliptical nebula shows a faint ansa, displaced 
about 0\farcs35 from the main nebula along the major axis to the SSE.  The 
feature is apparent in the clear image and in the [\ion{N}{2}] lines, but 
not in H$\alpha$ or [\ion{O}{3}]. The nebula is very bright, with a central 
peak of emission in H$\alpha$ that is absent in [\ion{N}{2}]. It is not 
clear if the CS is detected above the nebular emission, but it is saturated 
in the broad-band image. 
(See Fig.~\ref{Neb_8}.)

{\it LMC--SMP~48.---}This elliptical nebula is relatively compact with 
emission that is fairly strongly peaked, except in the emission of 
[\ion{N}{2}] and other low-ionization lines, where the flux distribution 
is somewhat more flat. The CS is probably not detected above 
the nebular emission. 
(See Fig.~\ref{Neb_8}.)

{\it LMC--SMP~49.---}This round nebula has a central cavity of reduced 
emission that is evident in all the emission lines. The intensity profile 
falls off more slowly than most other nebulae in this sample. The enhanced 
emission in the NE limb might indicate an interaction with gas outside 
of the illuminated shell. A faint CS is detected in the broad-band image.  
(See Fig.~\ref{Neb_9}.)

{\it LMC--SMP~51.---}This nebula is probably not resolved. The [\ion{O}{3}] 
is quite strong, and [\ion{N}{2}] is present but weak. 
(See Fig.~\ref{Neb_9}.)

{\it LMC--SMP~52.---}This somewhat strange nebula has an elliptical outer 
contour, but some inner structure. In the light of the faint [\ion{N}{2}] 
lines there appears to be an outer ring of emission, though in [\ion{O}{3}] 
there appears to be in addition a low-contrast, inner band or waist that 
extends from the ESE limb to the WNW. The CS is probably not detected.  
(See Fig.~\ref{Neb_9}.)

{\it LMC--SMP~57.---}This nebula has a round outer contour to what may be 
an attached outer shell. The inner shell is very elliptical, and like the 
bright CS, is displaced from the center of the outer shell by 0\farcs096 
arcsec (0.023 pc) to the ENE. Interestingly, this nebula has a plume (or 
jet?) of emission extending in the opposite direction (to the WSW) about 
2\farcs6 (0.13 pc) from the CS. Although the plume is evident in both 
[\ion{O}{3}] and H$\alpha$, it is about one tenth the surface brightness 
of the main shell and is therefore not visible in the figure. 
(See Fig.~\ref{Neb_9}.)

{\it LMC--SMP~61.---}The emission from this elliptical nebula is relatively 
uniform in the center in the broad-band and the high-ionization emission 
lines. There is some enhanced emission on the SE and NW limbs in the 
light of [\ion{N}{2}] and the other low-ionization lines, though not quite 
enough for the nebula to be classified as BC. The CS is easily visible in 
the broad-band image. 
(See Fig.~\ref{Neb_9}.)

{\it LMC--SMP~62.---}This elliptical nebula is unusual in that the 
innermost contours resemble a bar or highly oblate ellipse with the major 
axis oriented roughly {\it perpendicular} to the major axis of the 
outermost contours. The image in low ionization lines such as 
[\ion{N}{2}] show that the the central bar is terminated on each end by 
a small region of enhanced emission, and that the outer nebula resembles 
more of a box or barrel, much like a bipolar nebula. Hence, the designation 
as E(bc). No CS is evident within the bright nebulosity in the broad-band 
image. 
(See Fig.~\ref{Neb_10}.)

{\it LMC--SMP~63.---}The emission from this elliptical nebula is strongly 
peaked in high ionization lines such as H$\alpha$, \ion{He}{1} $\lambda$6678, 
and [\ion{O}{3}], but the emission in the low-ionization lines such as 
[\ion{N}{2}], [\ion{S}{2}], and [\ion{O}{1}] shows a ring structure, with 
enhanced brightness on the SW edge. The CS is probably visible, in spite 
of the intense and strongly peaked nebular emission. 
(See Fig.~\ref{Neb_10}.)

{\it LMC--SMP~64.---}This nebula is unresolved but very bright. The 
spectrum is clearly that of a PN, but the H$\alpha$ and H$\beta$ emissions 
are slightly broader than the other nebular lines, and the profiles are 
asymmetric, with enhanced emission to the red of the line centroid. There 
are many other faint emission lines present in the spectrogram, including 
\ion{He}{1} and \ion{He}{2} that could be of either nebular or stellar 
origin. A central star is visible, though saturated, in the broad-band 
image. A faint band structure can be seen in the G750M continuum, redward 
of about 6700~\AA. Given the complex stellar features and the apparent 
brightness of the continuum, it is possible that the the central ionizing 
source has a binary companion. A deep, high-resolution spectrum should be 
obtained to study the full spectrum in more detail. 
(See Fig.~\ref{Neb_10}.)

{\it LMC--SMP~67.---}This bipolar nebula shows in the broad-band image 
asymmetric wings eminating from a pinched waist, surrounding a bright CS. 
The detailed morphology in the individual nebular lines is very similar 
to LMC--SMP~30 (see Paper~I), in that the emission shows marked 
ionization stratification: i.e., the emission from [\ion{O}{3}] tends to be 
more spatially confined and shows less structure than H$\alpha$, while 
the low ionization lines of [\ion{N}{2}] show interlocking rings of 
emission surrounding a central cavity of lower emission. 
(See Fig.~\ref{Neb_10}.)

{\it LMC--SMP~68.---}The emission in this elliptical nebula is dominated 
by bright knots of emission on the N, W, and S sides, though fainter knots 
extend slightly from the outer elliptical contour on the E and W limbs. 
The knots are even more sharply defined in the [\ion{O}{3}] image, but 
the H$\alpha$ image shows only a moderate brightening on the W side. 
There is no emission in [\ion{N}{2}] or any other low ionization line. 
The CS is easily visible in the broad-band image. 
(See Fig.~\ref{Neb_10}.)

{\it LMC--SMP~69.---}This bipolar nebula has a complex morphology in 
detail. The most prominent feature is a central (projected) ring that 
defines the ``waist" (as is seen in LMC--Sa~107, Mo~21, and SMP~91), with 
embedded bright emission knots. Extending outward at least 1\farcs8 
from the waist are filiments of emission. Once again, the [\ion{O}{3}] and 
H$\alpha$ emission is spatially more confined than in [\ion{N}{2}], where 
a central cavity is evident, suggesting some degree of ionization 
stratification. The emission in [\ion{N}{2}] is much brighter than 
H$\alpha$, such that F($\lambda$6548+$\lambda$6583)/F($\lambda$6563) 
probably exceeds 3.  The [\ion{N}{2}] image shows more complicated 
structure in the ring, and faint emission extending away from the waist 
at least 2\farcs3, and surrounding a central cavity of reduced emission. 
The CS is not evident in the broad-band image, although an apparent field 
star just to the S of the center might be a CS candidate. 
(See Fig.~\ref{Neb_11}.)

{\it LMC--SMP~73.---}This elliptical nebula has a bipolar core that is 
evident in all emission lines, although the greater contrast between the 
nebular core and the lobes in the [\ion{N}{2}] emission suggest some 
ionization stratification in the nebula. No CS is evident in the 
broad-band image. 
(See Fig.~\ref{Neb_11}.)

{\it LMC--SMP~74.---}This elliptical nebula has very prominent bipolar 
characteristics, and might have been classified bipolar were it not for 
the nearly perfect elliptical contours at the level of a few percent of 
the peak emission. The central bar of emission bisects what looks to 
be two loops of emission along the major axis. Two bright, short arcs of 
emission eminate from the ends of the bar, in opposite directions and 
perpendicular to the bar, and then fade into the loops. 
The CS is marginally detected above the bright nebular emission in the 
broad-band image. 
(See Fig.~\ref{Neb_11}.)

{\it LMC--SMP~75.---}This nebula is barely resolved, making the morphological 
classification uncertain. The low ionization lines, including [\ion{N}{2}], 
have a flatter peak profile than the broad-band and [\ion{O}{3}] images. 
The CS may have been detected, but is saturated in the broad-band image.  
(See Fig.~\ref{Neb_11}.)

{\it LMC--SMP~82.---}This small, elliptical nebula has a very bright  
core of emission, particularly in the light of [\ion{O}{3}]. The larger 
and relatively flat-topped profile in [\ion{N}{2}] and other low-ionization 
lines reveals significant ionization stratification. The H$\alpha$ image 
shows an extension of emission to the blue of about 3.6~\AA\ (or if the 
extension is spatial, to the SSW by 0\farcs3). This extension is not seen 
in any other detected emission line, so it is not clear whether this 
feature is genuine. The CS is detected in the broad-band image. 
(See Fig.~\ref{Neb_11}.)

{\it LMC--SMP~83.---}This nebula has a very complex structure, seen in 
all emission lines, and is probably bipolar (though it looks somewhat  
point-symmetric). The pinched waist is over 
2\farcs0, but the largest of the four main loops of emission extends at 
least 3\farcs4 (0.84 pc) from the center. An extensive spectroscopic study 
of the dynamics of this nebula was published by \citet{Pena_etal2004}. 
The CS is easily detected in the broad-band image. 
(See Fig.~\ref{Neb_12}.)

{\it LMC--SMP~84.---}This elliptical nebula has rather weak [\ion{N}{2}], 
but unlike the emission in [\ion{O}{3}] or the recombination 
lines, the emission is confined to narrow columns along the major axis 
of the nebula, indicating an interesting ionization structure. The CS 
is likely detected above the nebular emission in the broad-band image. 
(See Fig.~\ref{Neb_13}.)

{\it LMC--SMP~88.---}This nebula is elliptical, and appears to have a 
faint extension surrounding a brighter core. The ionization appears to 
be somewhat stratified, judging by the larger and more pronounced ring 
of emission in [\ion{N}{2}] than in [\ion{O}{3}]. The cavity within the 
ring appears to be filled in H$\alpha$ with extended wings of probable  
stellar emission. The CS is easily visible in the broad-band image. 
(See Fig.~\ref{Neb_13}.)

{\it LMC--SMP~89.---}This elliptical nebula shows some hint of bipolar 
structure in the [\ion{O}{3}] image, with somewhat box-like inner contours. 
The bipolarity is less 
obvious in H$\alpha$, and is not present at all in the [\ion{N}{2}] 
emission. The low ionization lines show an emision deficit in the center 
of an irregular ring, revealing some ionization stratification in the 
nebula. There is no CS evident in the broad-band image. In many ways 
this object is similar to, though smaller than, LMC--SMP~62. 
(See Fig.~\ref{Neb_13}.)

{\it LMC--SMP~91.---}This bipolar nebula shows a classic butterfly shape 
in the broad-band, with a pinched waist of roughly 1\farcs75 that bisects 
what appears to be a ring of emission in projection.  
The emission in [\ion{N}{2}] is much brighter than H$\alpha$, such that 
F($\lambda$6548+$\lambda$6583)/F($\lambda$6563) probably exceeds 5.  The 
[\ion{N}{2}] image shows more complicated structure in the ring, and 
faint emission extending away from the waist at least 2\farcs3, and 
surrounding a central cavity of reduced emission. The cavity is far less 
apparent in the H$\alpha$ image. The nebula is fairly weak in [\ion{O}{3}] 
but the ring is very apparent. The CS is not evident in the broad-band image. 
(See Fig.~\ref{Neb_12}.)

{\it LMC--SMP~92.---}The emission in the E lobe of the bipolar core of 
this elliptical nebula is roughly 10\% stronger than the W lobe in the 
light of H$\alpha$ and [\ion{O}{3}], but the relative strengths are 
about equal in the light of [\ion{N}{2}]. No CS is evident in the 
broad-band image. 
(See Fig.~\ref{Neb_13}.)

{\it LMC--SMP~98.---}This small, round nebula has a relatively flat-topped 
spatial profile, though a small deficit of emission is apparent in 
[\ion{N}{2}] and the other, low-ionization lines, revealing some ionization 
stratification within the nebula. The CS is probably not detected.  
(See Fig.~\ref{Neb_13}.)

{\it LMC--SMP~101.---}This slightly elliptical nebula has small, faint 
ansae extending from the NNE and SSW limbs. The main nebula is a mostly 
filled shell, terminating in a ring at the outer boundary. Interestingly, 
the ansae are not symmetric about the center, but are displaced slightly, 
and lie roughly along a darker band that passes just to the E of the 
bright CS. The nebular features are similar in all of the emission lines, 
though the [\ion{O}{3}] image shows somewhat more detailed, if disorganized, 
structure. 
(See Fig.~\ref{Neb_14}.)

{\it SMC--JD~12.---}This nebula is probably bipolar, judging by the nearly 
orthogonal orientation between the brighter inner core and the outer regions.
This nebula is very apparent in the ACS/WFC F555W band, detected but faint 
in the F814W band, and not detected in the F435W band. No CS is evident 
in the any of the broad-band images. 
(See Fig.~\ref{Neb_16}.)

{\it SMC--SMP~2.---}This round nebula has an asymmetric brightness profile 
in broad-band light and in the high-ionization lines of H$\alpha$ and 
[\ion{O}{3}], with higher emission on the S side. The low ionization lines 
such as [\ion{N}{2}] show emission in an elliptical ring, again with 
somewhat brighter emission on the S edge. No CS is evident in the 
broad-band image. 
(See Fig.~\ref{Neb_15}.)

{\it SMC--SMP~3.---}This elliptical nebula has a very distinct bipolar 
core. Interestingly, only the lobes can be seen in the [\ion{N}{2}] 
emission. The CS is very apparent in the broad-band image. 
(See Fig.~\ref{Neb_15}.)

{\it SMC--SMP~15.---}The emission from this nebula is strongly peaked in 
both the broad-band image, as well as the high ionization lines of H$\alpha$ 
and [\ion{O}{3}]. The emission in [\ion{N}{2}] is not as strongly peaked, 
and from the $\lambda$6583 line it is possible to make an improved estimate 
of the nebular size, which is slightly broader than a point source. 
The CS might have been detected had it not saturated in the broad-band image. 
(See Fig.~\ref{Neb_15}.)

{\it SMC--SMP~16.---}This bright, compact, slightly elliptical nebula shows 
strongly peaked emission in the broad-band image and the high-ionization 
lines of  H$\alpha$ and [\ion{O}{3}]. The emission in [\ion{N}{2}] is 
somewhat more box-shaped, with a hint of enhanced brightness on the E and W 
limbs. The CS is probably detected in the broad-band image. 
(See Fig.~\ref{Neb_15}.)

{\it SMC--SMP~28.---}This bright, compact nebula is round but appears to 
have a faint, diffuse tail of emission extending about 0\farcs9 to the NW 
from the central star. No G750M spectrum is available. 
(See Fig.~\ref{Neb_15}.)

\section {Discussion}

\subsection {Broad Characteristics}

\begin {figure}
\epsscale{0.5}
\plotone {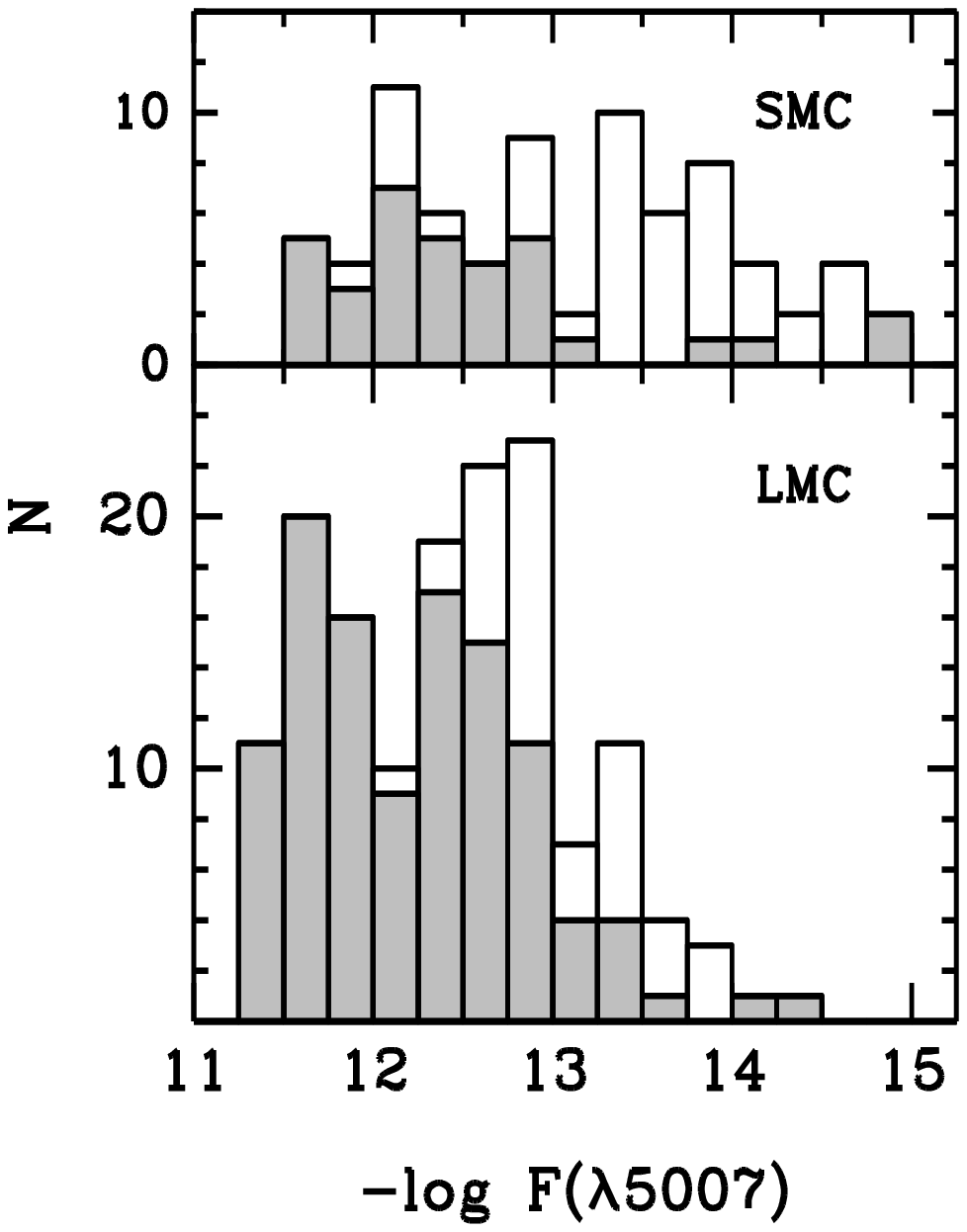}
\figcaption[f20.eps]{ 
Histogram of PN with published fluxes in [\ion{O}{3}] $\lambda$5007 
(erg cm$^{-2}$ s$^{-1}$) for the 
SMC ({\it upper}) and LMC ({\it lower}). Nebulae that have been observed 
with {\it HST} are indicated ({\it shaded}) in each histogram.
\label{Completeness}}
\end {figure}

As discussed in \S1, there are now 114 PNe in the LMC and 35 PNe in the SMC 
for which {\it HST} images are known to exist. Owing to the nature of the surveys, 
these samples are heavily weighted toward the brighter nebulae. This is 
illustrated in Figure~\ref{Completeness}, which shows the distribution of 
all MCPN with measured fluxes in [\ion{O}{3}] $\lambda$5007; objects that 
have also been imaged with {\it HST} are indicated in the shaded 
portions of the histograms. 
The markedly different relative numbers of known, faint PNe between the LMC 
and the SMC is almost certainly due to the superior completeness of the 
SMC sample \citep{JD02}. 
Of all confirmed LMC PNe, only about 40\% have published absolute fluxes 
in [\ion{O}{3}] $\lambda$5007; in the SMC the fraction is closer to 70\%. 
More data for the LMC should be available once ongoing surveys are 
published in full \citep[e.g.][]{ReidParker06}. \citet{JD02} published a 
uniform and deep survey of a significant portion of the SMC, so that this 
flux-complete sample can be studied in considerable detail. 

\begin {figure}
\epsscale{0.5}
\plotone {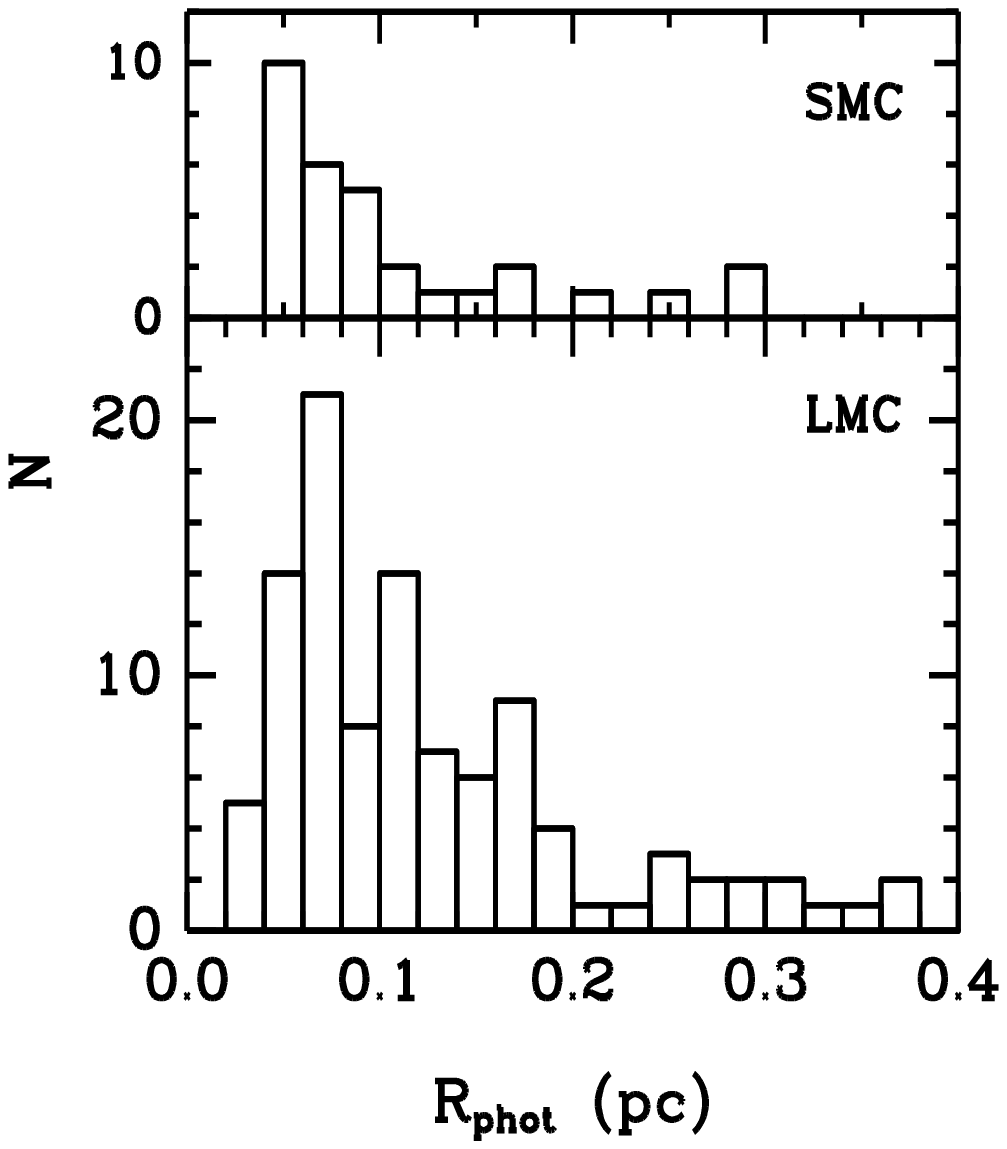}
\figcaption[f21.eps]{
Histogram of nebular photometric radii for the SMC ({\it upper}) 
and LMC ({\it lower}). 
\label{Rad_hist}}
\end {figure}

\begin {figure}
\epsscale{0.6}
\includegraphics[angle=-90]{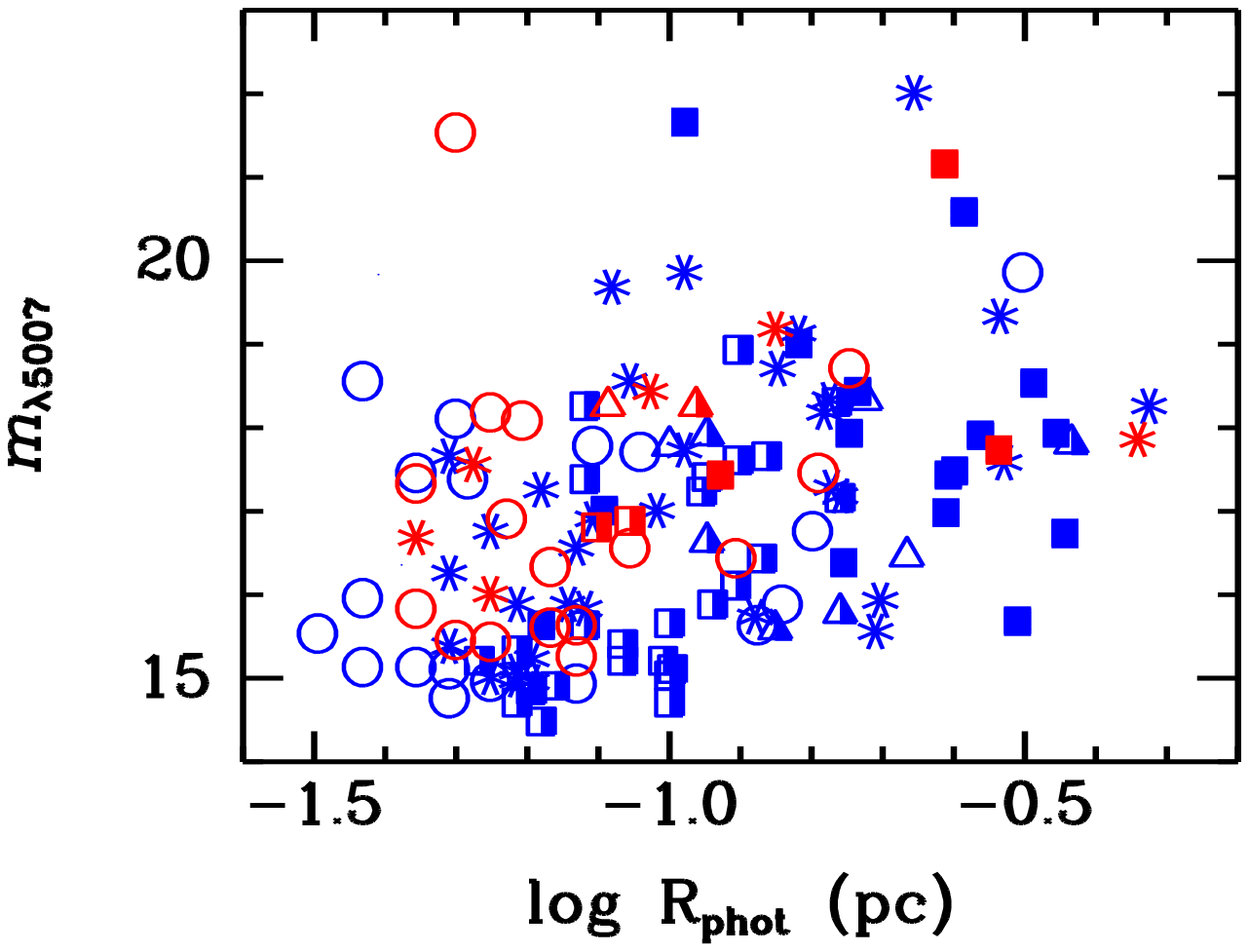}
\figcaption[f22_color.eps]{ 
Variation of the flux in [\ion{O}{3}] $\lambda$5007, uncorrected for 
interstellar extinction and expressed in magnitudes 
($m_{5007} = -2.5 \log F_{5007} - 13.74$), with photometric 
radius for the LMC and SMC 
nebulae. Symbols indicate morphological type: round ({\it circles}), 
elliptical ({\it asterisks}), quadrupolar ({\it half-filled triangles}), 
bipolar ({\it filled squares}), bipolar core ({\it half-filled squares}), 
and point-symmetric ({\it open triangles}).  
The electronic version of this paper shows LMC PNe in blue symbols, and 
SMC PNe in red. 
\label{Flux_rad}}
\end {figure}

The bulk of MCPNe in the {\it HST} samples are small, with a median 
photometric radius of 0.10 pc in the LMC and 0.08 pc in the SMC, as shown 
in Figure~\ref{Rad_hist}. Interestingly, there is only a rough correlation 
between the magnitude in [\ion{O}{3}] $\lambda$5007 \citep[$m_{5007}$, 
following][]{Ciard_etal89} and R$_{phot}$, in the sense that fainter PNe 
tend to be larger. As 
illustrated in Figure~\ref{Flux_rad}, successive bins in the [\ion{O}{3}] 
planetary nebula luminosity function (PNLF) sample a wide range of 
nebular sizes and morphological types, except for the brightest two 
magnitude bins where the PNe are typically small and bipolar nebulae 
are significantly under-represented. Although this {\it HST} sample is 
seriously incomplete at faint flux levels, the relative number of bipolars 
is much larger for nebulae fainter than about 2 mag below the brightest. We 
speculate that the robustness of the PNLF accross galaxy type may be in part 
due to the heterogenerity of nebular sizes, types, and chemical enrichments 
(see below) within all but the brightest one or two magnitude bins. For 
PNe within one magnitude of the brightest it is interesting to note that all 
are well resolved, and the median radius of $\sim$0.06 pc suggests that most 
MCPNe do not attain their maximum [\ion{O}{3}] brightness for some time 
after the material in the main shell is ejected by the CS. These properties, 
and the presence of only one bipolar PN (in the LMC) among these nebulae, 
suggests that few if any of the most luminous MCPNe are composed of a 
post-asymptotic giant branch star and a close accreting white dwarf 
companion, as asserted by \citet{Soker_06}. 
In this discussion of the PNLF, no correction has been applied for the 
generally modest amount of extinction, which is usually attributed to 
foreground dust. The median value for $c$, the logarithmic extinction at 
H$\beta$, is only 0.19 for the LMC sample and 0.10 for the SMC. The 
amount of optical extinction does not appear to correlate with morphological 
type or size. 

\subsection {Morphology and Evolution}

\begin {figure}
\epsscale{0.6}
\includegraphics[angle=-90]{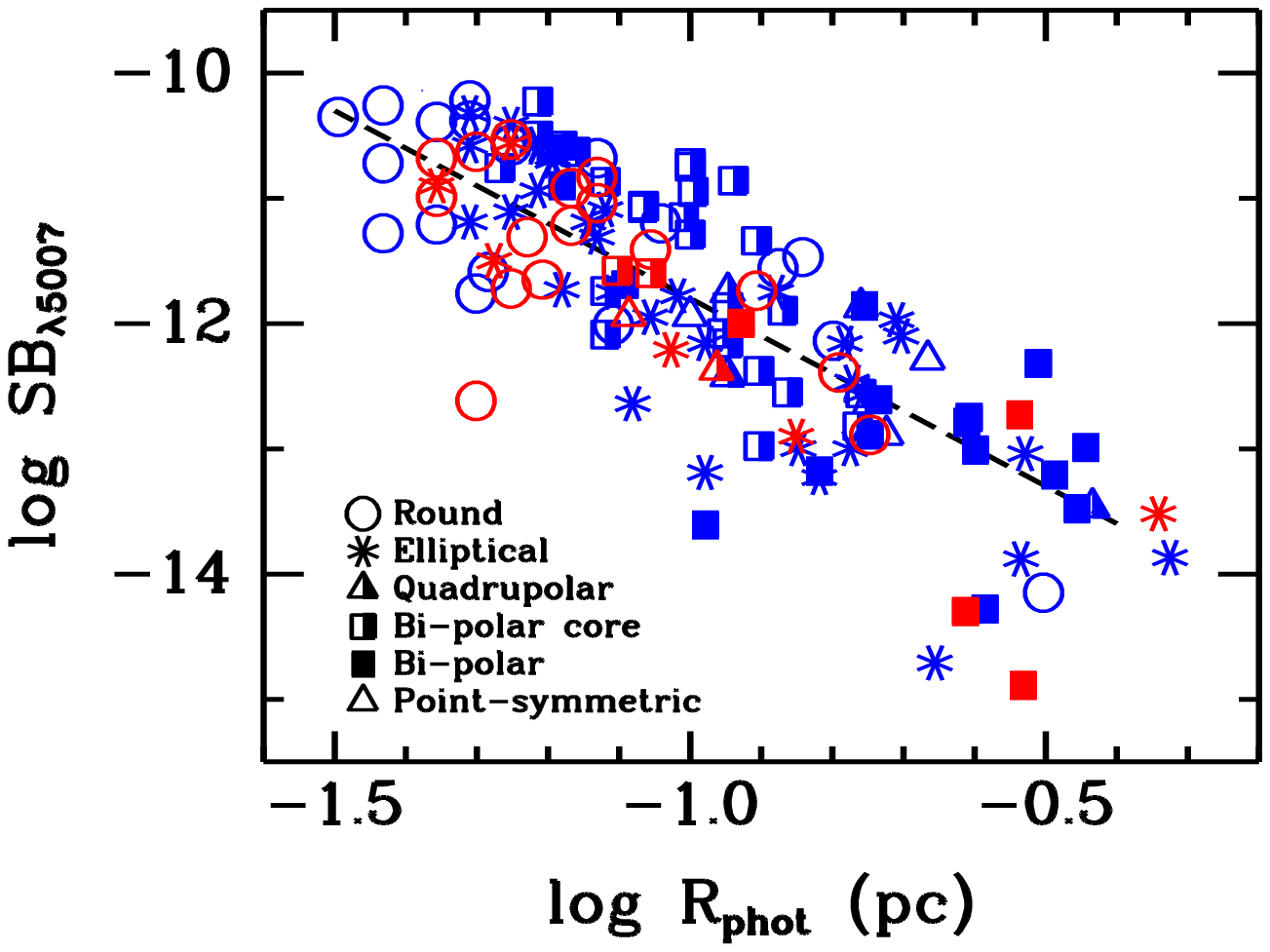}
\figcaption[f23_color.eps]{ 
Decline of extinction-corrected, average surface brightness in [\ion{O}{3}] 
$\lambda5007$ (SB$_{[O~III]}$, in erg cm$^{-2}$ s$^{-1}$ arcsec$^{-2}$) 
with R$_{phot}$ is consistent with an 
R$^{-3}$ power law ({\it dotted line}). The various morphological types 
are represented by different symbols, as in Fig.~\ref{Flux_rad} and in 
the legend. 
The electronic version of this paper shows LMC PNe in blue symbols, 
and SMC PNe in red. 
\label{SB_rad}}
\end {figure}

Paper~I presented a preliminary discussion of the decline in nebular 
surface brightness with photometric radius. The discussion focussed on the 
possibility of an evolution in nebular morphological type with age, and 
suggested that understanding the complexity of the co-evolution in the 
nebula and its central star is a prerequisite for fully understanding the 
diagram. We revisit this plot in Figure~\ref{SB_rad}, but update it to show 
the extinction-corrected, average surface brightness 
(in erg cm$^{-2}$ s$^{-1}$ arcsec$^{-2}$) as a function of the physical 
size of the ionized shell (in pc) for all LMC and SMC PNe in the {\it HST} 
sample. To compute the average surface brightness we divided our 
[\ion{O}{3}] $\lambda$5007 flux, corrected for extinction (using our 
measurements of the Balmer decrement where possible: see \S~2), by the 
area of a circle with radius equal to our photometric radius. 
The morphological types are indicated symbolically. The trend of the 
data along $R^{-3}$ is quite apparent, as are the similarity of slope, 
normalization, and dispersion about the trend line ($\sim0.75$ dex) for 
both LMC and SMC PNe. It is remarkable that the surface brightnesses 
of a large ensemble of PNe can be so simply characterized. 
\cite{Renzini_81} predicted a rather different form for the decline of PN 
surface brightness with radius. Based on the simple argument that for 
low density gas the volume emissivity is proportional to the product of 
the densities of electrons and ions, he argued that the volume emissivity 
declines as $R^{-6}$ for an expanding nebula of fixed mass, so the average 
surface brightness of the projected nebula 
should decline as $R^{-5}$. Real nebulae are much more complex than 
this however, and other important factors can have a large impact on the 
evolution of the emissivity. These factors include compression of the gas 
as the fast CS wind impinges on the inner boundary of the main shell, the 
weak shock associated with the passage of an ionization front through 
neutral gas, and the progressive ionization of greater amounts of the 
surrounding neutral gas etc. \citep[e.g.,][]{Vill_etal02, Perinotto_etal04}. 
It is important to note that, for nebulae with high optical depth, it is the 
radius of the O$^{2+}$ zone, not the radius of the mass-bounded shell, 
that is being measured. Even for optically thin nebulae, R$_{phot}$ would 
be defined more by the details of the density profile than by some physical 
boundary, and in any case it is not necessarily linearly related to the 
kinematical age of the nebula. Thus the surface brightness relation may 
be more closely connected to characteristics of the evolution of 
the PN ionization than to the expansion of the shell per se. 

In spite of the simplicity of the surface brightness relation, it is probably 
not correct to conclude that the trajectory of an individual PN is at all 
times parallel to the trend line. Several of the nebulae in Fig.~\ref{SB_rad} 
lie significantly below the trend, possibly indicating a relatively low 
ionization parameter. Many of these nebulae, particularly those with radii 
larger than about log~R$_{phot} = -0.7$, are bipolar. 
PN evolutionary models by \citet{Vill_etal02} and \citet{Perinotto_etal04} 
predict that the most massive CSs (which have also lost the most mass 
on the AGB) will never fully ionize their surrounding gas shells, and the 
PN will never become optically thin because the CS evolves so quickly to 
low luminosity. Exactly when during its evolution a PN becomes optically 
thin depends on a number of factors, and it has a significant bearing on 
its detectability. 

\begin {figure}
\epsscale{0.6}
\includegraphics[angle=-90]{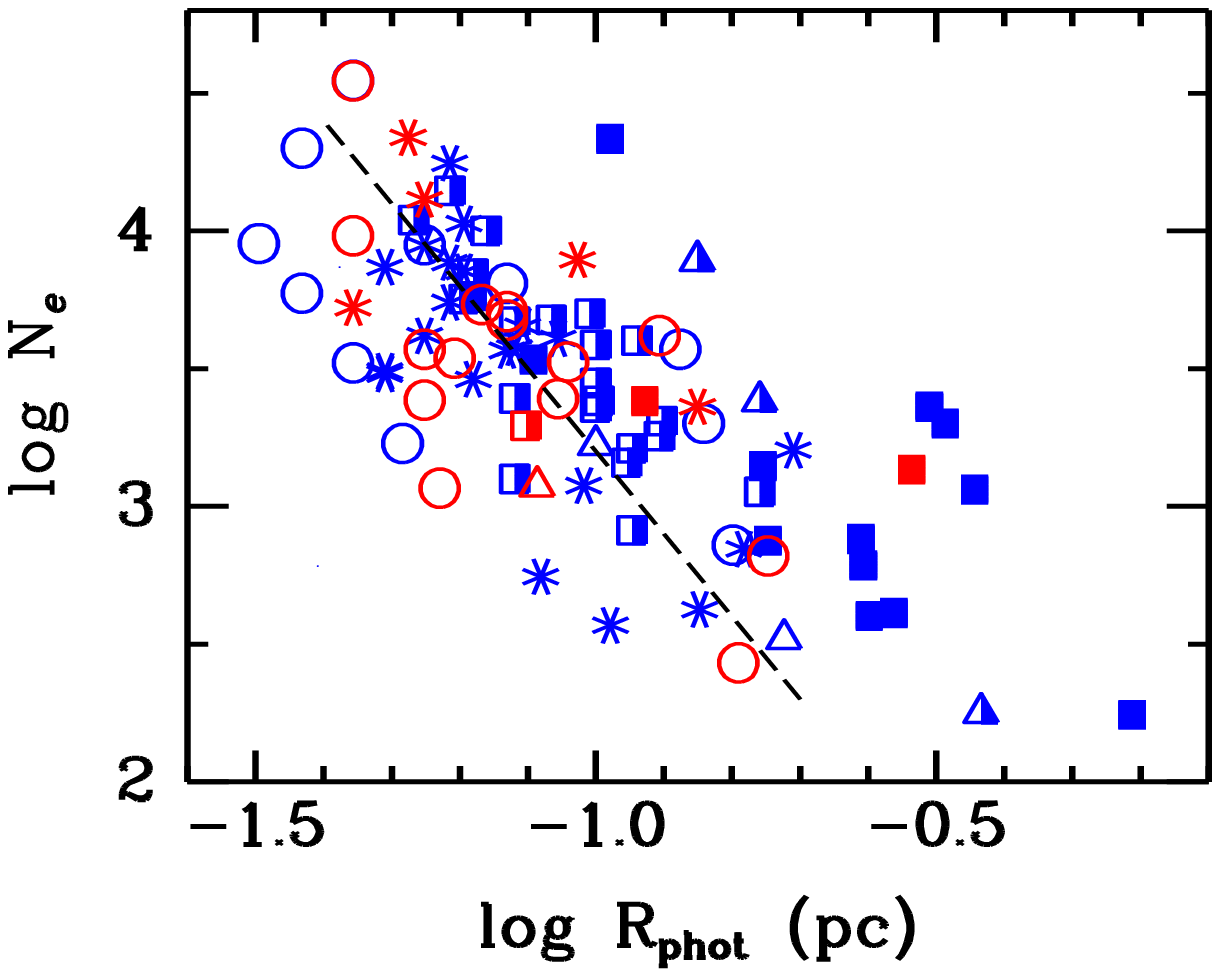}
\figcaption[f24_color.eps]{ 
Nebular electron density vs. nebular (photometric) radius in pc. The nebular 
morphological types are represented with symbols as in Fig.~\ref{Flux_rad}. 
Dashed line shows an example decline in N$_e$ as $R^{-3}$. 
The electronic version of this paper shows LMC PNe in blue symbols, and 
SMC PNe in red.
\label{Dens_Rad}}
\end {figure}

Further 
insight into the optical depth of PNe in this sample can be drawn from 
Figure~\ref{Dens_Rad}, where the electron density is plotted against 
R$_{phot}$. For this plot we have derived nebular densities from the 
ratio of the [\ion{S}{2}] $\lambda$6723 doublets presented in this paper, 
and in Papers~II and III when available, assuming an electron temperature 
of 10,000 K. Other densities were taken from \citet{MeathDop91a, 
MeathDop91b}: from [\ion{S}{2}] where possible and from [\ion{O}{2}] 
$\lambda$3727 if not. Densities for a few additional objects were 
derived from our own unpublished, ground-based optical spectra. Basic 
density information is available for most, but not all nebulae in this study; 
there is clearly a need for additional, deep spectroscopy for PNe in the 
Magellanic Clouds. 
It should be noted that densities larger than log~N$_e$ of 3.7 are less 
accurate, as this is well above the criticial densities for the 
[\ion{S}{2}] $\lambda\lambda$6716, 6731 lines. 
The figure shows a very clear decline in density with increasing radius, 
roughly following a $R^{-3}$ power law. However, for radii larger than 
log~$R_{phot}\sim-0.7$ there are many nebulae, all of which 
are bipolar (or similar morphology), that lie substantially above the 
trend defined by the smaller ionized shells. Indeed, nearly all of the 
bipolar nebulae have higher density at a given radius than those of other 
morphological types, and a few of them also lie substantially below the 
surface brightness trend line in Fig.~\ref{SB_rad} for log~$R_{phot}>-0.7$. 

One might naively expect that the density of an individual nebula with 
constant ionized mass would decline as $R^{-3}$ as the shell expands, 
but as mentioned above real nebulae are more complex. For instance, 
the effects of compression on the inner PN boundary by the fast wind 
and radiation from the CS would affect the trajectory, as would the initial 
density contrast between equitorial and polar regions of the proto-PN 
\citep{ZhangKwok98} when the PN shell was formed. 
However, the mean density as measured by the [\ion{S}{2}] lines 
may not be representative of all the gas. That is, the [\ion{S}{2}] emissivities 
will be strongest near the critical density (1.4--3.6$\times 10^3$ cm$^{-3}$), 
so that the measured density will to some degree be intensity weighted, 
and will depend upon the details of the ionization in the region where 
density can be measured with S$^+$. 
But the remarkable appearance of many relatively dense nebulae beyond 
log~$R_{phot} > -0.7$, some of which have modest ionized mass 
($\lesssim0.2$M$_{\sun}$), suggests that additional effects are in play, 
at least for bipolars. 
For nebulae with high optical depth, where the ionization front has not 
yet progressed through the entire extent of the nebular gas, S$^+$ is most 
likely to be detected close to the ionization front 
(owing to its low ionization potential), and therefore in a region of elevated 
density. For these reasons one might expect to measure high [\ion{S}{2}] 
density throughout most of the life of PNe with the most massive progenitor 
stars (i.e., those that never fully ionize the surrounding nebulae). Thus, it 
seems plausible that nebulae which remain optically thick would tend to 
lie above the $R^{-3}$ trend even at large radius. 
In short, we believe the broad properties of bipolar PNe presented here: 
surface brightnesses that lie on or below the $R^{-3}$ trend line, unusually 
high measured density at large radii, sometimes modest ionized mass, and 
their correlation (at least in the Galaxy) with the most massive PN 
progenitors provide support for the prediction that the most massive 
CSs remain optically thick throughout the PN phase. 
Note that conclusions such as these would not be possible from a sample 
of PNe for which the individual distances are not known accurately, as is 
typical of current Galactic samples. 

\subsection {Morphology and Chemical Enrichment}

Chemical enrichments in PNe, primarily in N and He, have been studied for 
decades, and their connection to more massive progenitors is perhaps 
most widely expressed as the Peimbert Type~I class \citep{Peimbert78}. 
The association of bipolar morphology in Galactic PNe with Type~I 
characteristics was first studied in detail by \citet{PTP83} and \citet{Kaler83}; 
Kaler noted that a disproportionate number of B nebulae had high values 
of N/O and elevated He/H, compared to nebulae with other morphologies. 
These findings lent support to the connection between B nebulae and more 
massive progenitors that experience hot-bottom burning, where C is largely 
converted to N and dredged-up 
in the progenitor AGB star prior to the ejection of the main PN shell. 
\Citet{Stang_etal00} studied the chemical enrichments in LMC PNe with known 
morphologies, and showed that bipolar and bipolar core morphology was 
correlated with elevated N and depressed C abundances outside of the 
Galactic context. 

\begin {figure}
\epsscale{0.6}
\includegraphics[angle=-90]{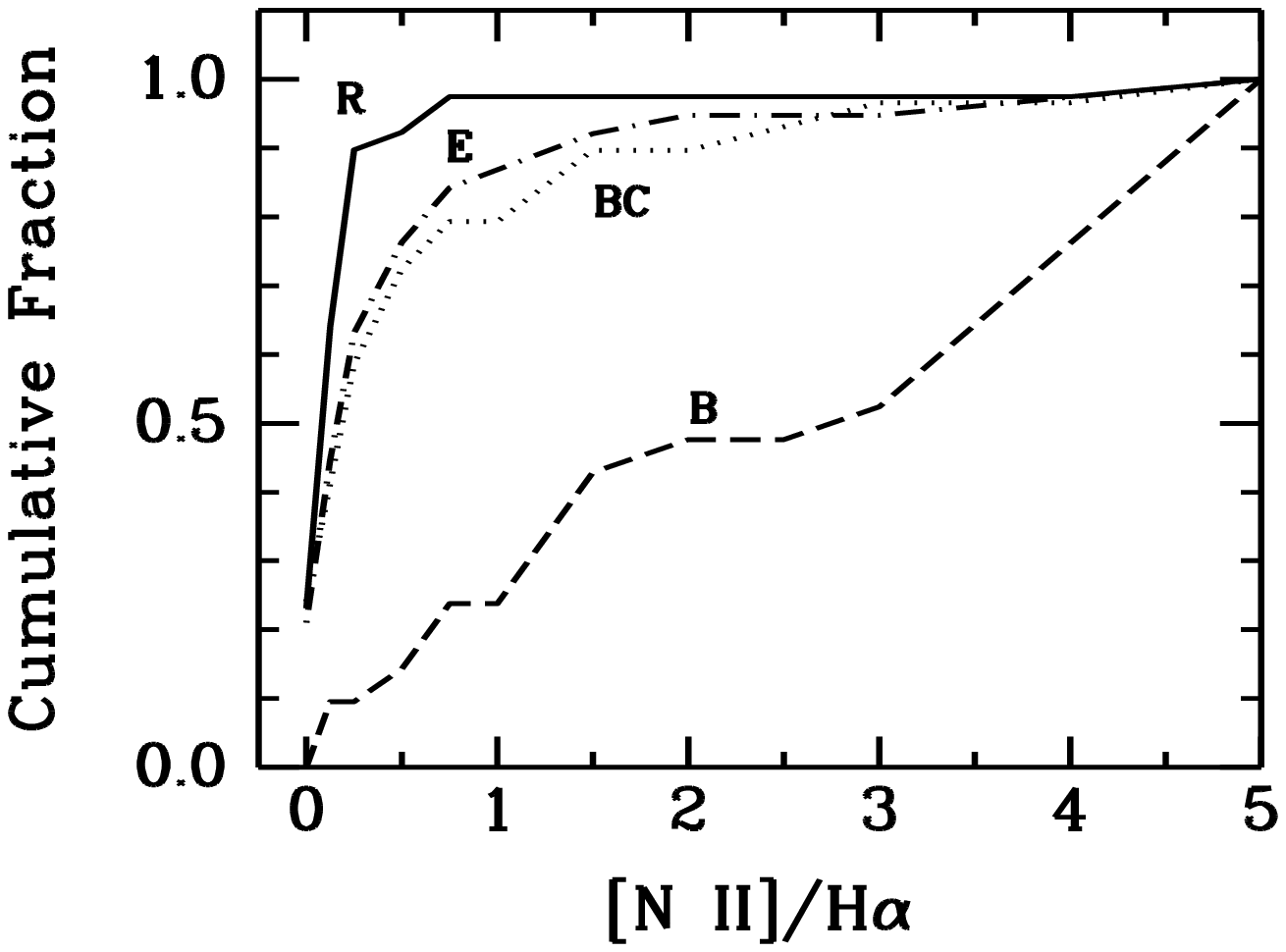}
\figcaption[f25.eps]{ 
Cumulative fraction of PNe as a function of the flux ratio 
F([\ion{N}{2}])/F(H$\alpha$) (see text) for each morphological type: 
Round ({\it solid}), Elliptical ({\it dot-dash}), Bi-polar Core 
({\it dotted}), and Bipolar ({\it dashed}). 
\label{N2_Ha}}
\end {figure}

\begin {figure}
\epsscale{0.6}
\includegraphics[angle=-90]{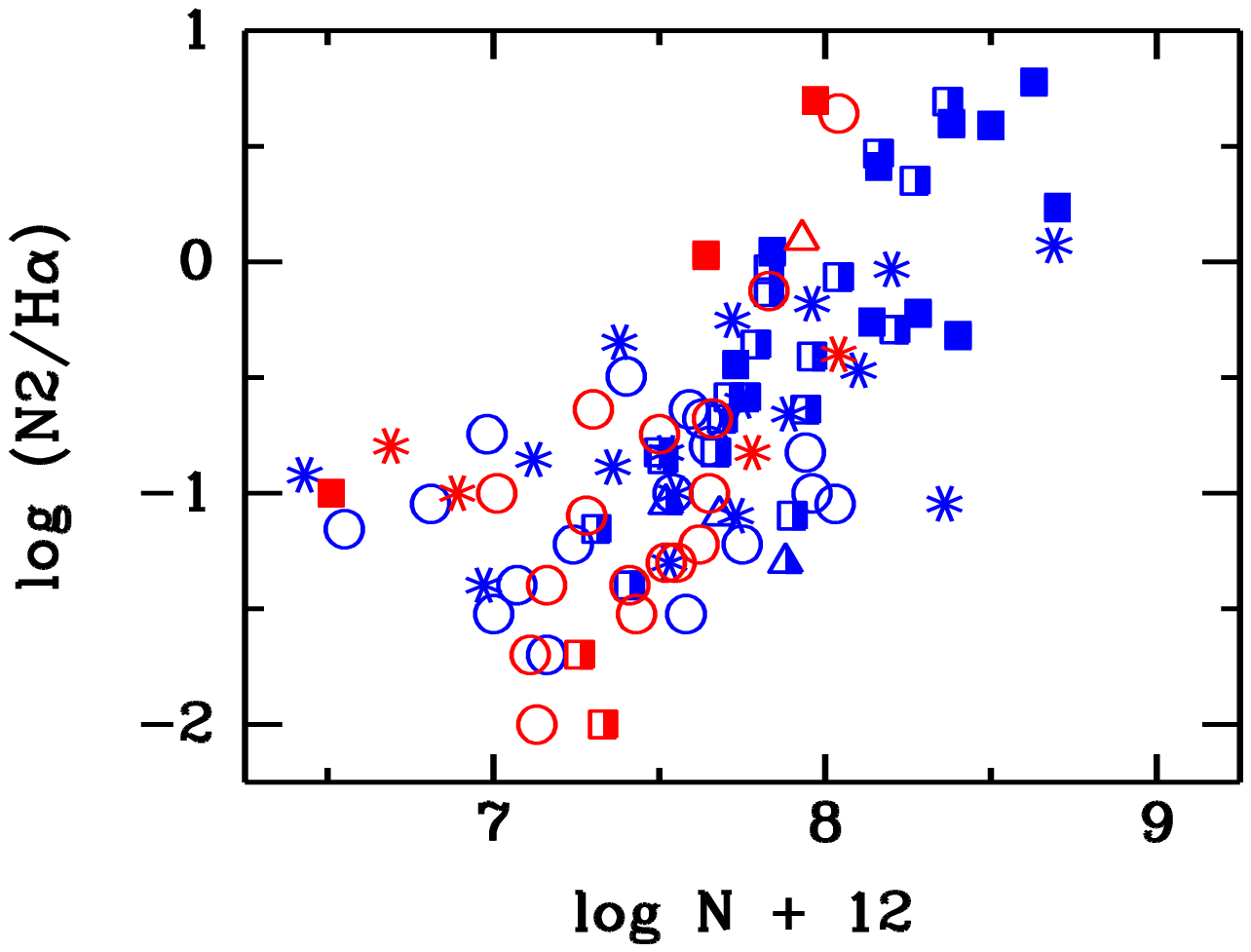}
\figcaption[f26_color.eps]{ 
F([\ion{N}{2}])/F(H$\alpha$) vs. N for the LMC and SMC nebulae. 
The N abundances were taken from \citet{SRM_98, LD_96}, and 
\citet{MeathDop91a, MeathDop91b}. 
Symbols indicate morphological type, as in Fig.~\ref{SB_rad}.
The electronic version of this paper shows LMC PNe in blue symbols, and 
SMC PNe in red.
\label{N2_N}}
\end {figure}

With our greatly enlarged sample of MCPN with known morphologies, plus 
measurements of the nebular [\ion{N}{2}] line fluxes, we can gain additional 
insight into the connection between morphology and chemical enhancements. 
Figure~\ref{N2_Ha} shows a cumulative histogram of the ratio of the fluxes 
in the [\ion{N}{2}] $\lambda\lambda$6548 + 6583 lines to that for H$\alpha$ 
(hereafter, N2/H$\alpha$) for four morphological types: R, E, BC, and B. 
The figure includes data for both LMC and SMC PNe, though if plotted 
separately the differences would be masked by small number statistics, 
since the SMC has fewer nebulae with known morphology and a lower 
fraction of bipolars. The [\ion{N}{2}] fluxes are from this paper and 
from Papers~II and III where available, or from \citet{MeathDop91a, 
MeathDop91b}. 
The plot shows that almost all R nebulae have N2/H$\alpha < 0.25$, and for 
all nebulae except one the ratio is less than 0.75. (The one exceptional 
R nebula is barely resolved, and the R classification is uncertain.) The 
histograms for the E and BC nebulae are nearly indistinguishable. 
Like the R nebulae about 20\% have N2/H$\alpha < 0.05$, but from there the 
curves rise more slowly: about 20\% of the nebulae have N2/H$\alpha > 1.0$, 
and for 10\% of the nebulae the ratio exceeds 1.5. But the most striking 
feature of the plot is the histogram for B nebulae, where none have a 
ratio near zero, and only 25\% have N2/H$\alpha < 1.0$. Indeed, most B 
nebulae have N2/H$\alpha > 3.0$. 
Although the N2/H$\alpha$ is only a proxy for N abundance, Figure~\ref{N2_N} 
shows that it is highly useful as a simple spectral indicator. 
A full abundance analysis of this MCPN sample is beyond the scope 
of this paper, but for the present purpose, nebulae with 
N2/H$\alpha >  1$ (the bulk of the B nebulae) almost certainly have  an 
elevated N abundance. Thus, Fig.~\ref{N2_Ha} provides perhaps the 
clearest evidence yet of the remarkable difference in chemical enhancements 
between PN morphological types. 

\section {Conclusions and Future Work}

We have presented images and slitless spectra of a final survey with STIS 
of MCPN, which when combined with prior work with {\it HST}, previous 
surveys that we have carried out, and with a few other, mostly 
serendipitous observations in the {\it HST} archive, yields the largest 
sample of resolved extragalactic PNe to date. The broad properties of 
the MCPN were described, as were the limits to our understanding that 
cannot be addressed further without high-resolution images of flux-complete 
samples. This paper also draws some important conclusions about the 
effects of optical depth in PNe, and sets the stage for future observations 
and modelling that will further our understanding of the evolution of 
planetary nebulae and their central stars. 
An analysis of the properties of the central stars is the subject of 
another paper \citep{Villa06}. There are a number of PNe in this sample 
for which detailed spectra are missing or incomplete. We have obtained 
spectra for some of them, from which we will derive essential phyisical 
information (such as densities), as well as chemical abundances. More 
deep, high-resolution images of PNe will be required in both the LMC and 
SMC to establish a large and complete sample to a photometric depth that 
is comparable to those of modern surveys (i.e., at least 6 mag of the 
[\ion{O}{3}] luminosity function), so that more firm conclusions can 
be drawn about the range of morphological, physical, and chemical 
properties of PNe. 

\acknowledgements 

We are grateful for the assistance of the very capable STScI staff in the 
preparation of this very extensive observing program, and to D.~Saga for 
the preparation of the images of the nebulae. We thank Drs. B.~Balick 
and J.~C.~Blades for their participation in the observing program. 
We are also grateful for many careful comments from the referee which 
helped to improve the presentation. 
Support for this work was provided by NASA through grants GO-09077 and 
GO-10251 from Space Telescope Science Institute, which is operated by the 
Association of Universities for Research in Astronomy, Incorporated, 
under NASA contract NAS5--26555.

\clearpage

%
\begin{deluxetable}{rlccrl}
\tabletypesize{\scriptsize}
\tablecolumns{6}
\tablewidth{0pc}
\tablecaption {HST Imaging Programs that Include Magellanic Cloud PNe 
\label{Surveys}}
\tablehead { 
\colhead {HST} & \colhead {}   & \colhead {}           & 
\colhead {}    & \colhead {No.} & \colhead {Reference} \\
\colhead {Program} & \colhead {PI} & \colhead {Instrument} & 
\colhead {Galaxy} & \colhead {Targets\tablenotemark{a}} & \colhead {Keys} \\
}
\startdata  
1046  & Blades, J.~C.   & WFPC     & LMC &  3 & 1 \\
1266  & Blades, J.~C.   & FOC      & LMC &  2 & 1 \\* 
      &                 &          & SMC &  2 &  \\ 
2266  & Dopita, M.      & WFPC     & LMC & 11 & 2 \\* 
      &                 &          & SMC &  4 &  \\ 
4075  & Blades, J.~C.   & FOC      & LMC &  5 & 5, 7, 8 \\* 
      &                 &          & SMC &  3 &  \\ 
4821  & Blades, J.~C.   & FOC      & LMC &  3 & 5, 8 \\* 
      &                 &          & SMC &  3 &  \\ 
4940  & Blades, J.~C.   & FOC      & LMC &  5 & 5, 8 \\* 
      &                 &          & SMC &  1 &  \\ 
5185  & Blades, J.~C.   & FOC      & LMC &  3 & 5, 8 \\
6407  & Doptia, M.      & WFPC2    & LMC & 12 & 4 \\ 
7303  & Hamann, W.-R.   & STIS:S\tablenotemark{b}   & LMC &  1 & 10 \\ 
7783  & Baum, S.        & STIS:I   & LMC &  1\tablenotemark{c} & 3 \\ 
8271  & Stangellini, L. & STIS:I,S & LMC & 29 & 4 \\ 
8059  & Casertano, S.   & WFPC2    & LMC & 1\tablenotemark{c} & \nodata \\ 
8663  & Stangellini, L. & STIS:I,S & SMC & 27 & 6 \\ 
8702  & Shaw, R.        & WFPC2    & SMC & 13 & 9 \\ 
9077  & Shaw, R.        & STIS:I,S & LMC & 54 & 11 \\ 
9285  & Baum, S.        & STIS:I   & LMC &  1\tablenotemark{c} & \nodata \\ 
9584  & Sparks, W.      & ACS:I    & LMC &  1\tablenotemark{c} & \nodata \\ 
9891  & Gilmore, G.     & ACS:I    & LMC &  1\tablenotemark{c} & \nodata \\ 
10126 & Olszewski, E.   & ACS:I    & SMC &  1\tablenotemark{c} & \nodata \\ 
10251 & Shaw, R.        & STIS:I,S & SMC & 5 & 11 \\ 
\enddata
\tablerefs{(1)\citet{Blades_etal92}, 
(2)\citet{Dopita_etal96}; (3)\citet{PlaitGull99}; (4)\citet{Shaw_etal2001}; 
(5)\citet{Stang_etal99}; (6)\citet{Stang_etal03}; (7)\citet{Vass96}; 
(8)\citet{Vass_etal98}; (9)\citet{Villa03}; (10)\citet{HeraldBianchi04}; 
(11)this paper .}
\tablenotetext{a}{Some PNe were observed in multiple programs.}
\tablenotetext{b}{Uncalibrated target acquisition image, to support 
spectroscopy.}
\tablenotetext{c}{Serendipitous discovery in pure parallel program image.}

\end{deluxetable} 

%
%
\begin{deluxetable}{lcllrc}
\tablecolumns{6}
\tabletypesize{\scriptsize}
\tablewidth{0pt}
\tablecaption {Observing Log \label{ObsLog}}
\tablehead {
\colhead {} & \colhead {} & \colhead {} & 
\colhead {} & \colhead {T$_{Exp}$} & \colhead {} \\
\colhead {Nebula} & \colhead {Date} & \colhead {Dataset} & 
\colhead {Disperser} & \colhead {(s)} & \colhead {N$_{Exp}$} 
}
\startdata  
\cutinhead{LMC Nebulae}
LMC--J~5     & 2001 Oct 10 & O6EL02010 & MIRVIS & 120 & 2 \\*
        &             & O6EL02020 &  G750M & 330 & 2 \\*  
        &             & O6EL02030 &  G430M & 660 & 2 \\
LMC--J~33    & 2002 Jan 22 & O6EL0L010 & MIRVIS &  300 & 2 \\*  
        &             & O6EL0L020 &  G750M & 1500 & 2 \\*  
        &             & O6EL0L030 &  G430M &  600 & 2 \\  
LMC--MG 4    & 2002 Apr 13 & O6EL0S010 & MIRVIS &  120 & 2 \\* 
        &             & O6EL0S020 &  G750M &  300 & 2 \\*
        &             & O6EL0S030 &  G750M & 1200 & 2 \\* 
        &             & O6EL0S040 &  G430M &  900 & 2 \\
LMC--MG 14   & 2002 Apr 30 & O6EL12010 & MIRVIS & 120 & 2 \\* 
        &             & O6EL12020 &  G750M & 480 & 2 \\*
        &             & O6EL12030 &  G430M & 600 & 2 \\
LMC--MG 16   & 2002 Feb 02 & O6EL14010 & MIRVIS &  120 & 2 \\*
        &             & O6EL14020 & G750M  &  300 & 2 \\*
        &             & O6EL14030 & G750M  & 1200 & 2 \\* 
        &             & O6EL14040 & G430M  &  600 & 2 \\  
LMC--MG 29   & 2002 Jan 29 & O6EL1H010 & MIRVIS &  120 & 2\\*
        &             & O6EL1H020 & G750M  & 1500 & 2 \\* 
        &             & O6EL1H030 & G430M  &  600 & 2 \\
LMC--MG 40   & 2001 Aug 14 & O6EL1S010 & MIRVIS &  120 & 2 \\*
        &             & O6EL1S020 &  G750M &  180 & 2 \\* 
        &             & O6EL1S030 &  G430M & 1000 & 2 \\  
LMC--MG 45   & 2001 Aug 19 & O6EL1X010 & MIRVIS &  120 & 2 \\*
        &             & O6EL1X020 &  G750M &  300 & 2 \\* 
        &             & O6EL1X030 &  G750M & 1200 & 2 \\*
        &             & O6EL1X040 &  G430M &  600 & 2 \\
LMC--MG 51   & 2001 Dec 10 & O6EL23010 & MIRVIS &  120 & 2 \\*
        &             & O6EL23020 &  G750M &  300 & 2 \\*  
        &             & O6EL23030 &  G750M & 1200 & 2 \\*
        &             & O6EL23040 &  G430M &  600 & 2 \\  
LMC--MG 70   & 2001 Aug 30 & O6EL2M010 & MIRVIS &  120 & 2 \\*
        &             & O6EL2M020 &  G750M &  300 & 2 \\*  
        &             & O6EL2M030 &  G750M & 1200 & 2 \\*
        &             & O6EL2M040 &  G430M &  600 & 2 \\  
LMC--Mo 7    & 2002 Feb 03 & O6EL39010 & MIRVIS &  120 & 2 \\*
        &             & O6EL39020 &  G750M &  300 & 2 \\*
        &             & O6EL39030 &  G750M & 1200 & 2 \\*  
        &             & O6EL39040 &  G430M &  600 & 2 \\  
LMC--Mo 21   & 2002 May 24 & O6EL3N010 & MIRVIS &  120 & 2 \\*
        &             & O6EL3N020 &  G750M &  300 & 2 \\*  
        &             & O6EL3N030 &  G750M & 1200 & 2 \\*  
        &             & O6EL3N040 &  G430M &  600 & 2 \\  
LMC--Mo 33   & 2002 Feb 28 & O6EL3Z010 & MIRVIS &  300 & 2 \\*
        &             & O6EL3Z020 &  G750M & 1620 & 2 \\*
        &             & O6EL3Z030 &  G430M &  480 & 2 \\
LMC--Mo 36   & 2001 Dec 11 & O6EL42010 & MIRVIS &  120 & 2 \\*
        &             & O6EL42020 &  G750M &  300 & 2 \\*  
        &             & O6EL42030 &  G750M & 1200 & 2 \\* 
        &             & O6EL42040 &  G430M &  600 & 2 \\  
LMC--Mo 47   & 2001 Aug 18 & O6EL4D010 & MIRVIS &  120 & 2 \\*
        &             & O6EL4D020 &  G750M &  300 & 2 \\*
        &             & O6EL4D030 &  G750M & 1200 & 2 \\*
        &             & O6EL4D040 &  G430M &  600 & 2 \\  
LMC--Sa 104a & 2002 Jun 29 & O6EL4M010 & MIRVIS &  120 & 2 \\*  
        &             & O6EL4M020 &  G750M &  300 & 2 \\* 
        &             & O6EL4M030 &  G750M & 1200 & 2 \\*
        &             & O6EL4M040 &  G430M &  600 & 2 \\  
LMC--Sa 107  & 2002 May 18 & O6EL4P010 & MIRVIS & 120 & 2 \\*
        &             & O6EL4P020 &  G750M & 640 & 2 \\*
        &             & O6EL4P030 &  G430M & 340 & 2 \\
LMC--Sa 117  & 2001 Aug 15 & O6EL4R010 & MIRVIS & 300 & 2 \\*  
        &             & O6EL4R020 &  G750M & 820 & 2 \\*
        &             & O6EL4R030 &  G430M & 640 & 2 \\  
LMC--Sa 121  & 2001 Aug 21 & O6EL4T010 & MIRVIS &  300 & 2 \\* 
        &             & O6EL4T020 &  G750M & 1500 & 2 \\*  
        &             & O6EL4T030 &  G430M &  600 & 2 \\  
LMC--SMP 3   & 2002 Apr 15 & O6EL4Y010 & MIRVIS & 120 & 2 \\*  
        &             & O6EL4Y020 &  G750M &  14 & 2 \\*  
        &             & O6EL4Y030 &  G750M &  70 & 2 \\*  
        &             & O6EL4Y040 &  G750M & 280 & 2 \\*  
        &             & O6EL4Y050 &  G430M &  60 & 2 \\* 
        &             & O6EL4YGMQ & MIRVIS &  15 & 1 \\  
LMC--SMP 5   & 2001 Sep 05 & O6EL4Z010 & MIRVIS & 120 & 2 \\*  
        &             & O6EL4Z020 &  G750M & 120 & 2 \\*
        &             & O6EL4Z030 &  G430M & 300 & 2 \\*  
        &             & O6EL4ZOJQ &  G750M & 600 & 1 \\  
LMC--SMP 6   & 2001 Aug 18 & O6EL50010 & MIRVIS & 120 & 2 \\*  
        &             & O6EL50020 &  G750M &  30 & 2 \\*  
        &             & O6EL50XBQ & MIRVIS &  15 & 1 \\*
        &             & O6EL50XCQ &  G750M &   1 & 1 \\*  
        &             & O6EL50XDQ &  G750M &   7 & 1 \\* 
        &             & O6EL50XGQ &  G430M &   2 & 1 \\  
LMC--SMP 11  & 2002 Jun 06 & O6EL52010 & MIRVIS & 300 & 2 \\*
        &             & O6EL52020 &  G750M & 450 & 2 \\*  
        &             & O6EL52030 &  G430M & 900 & 2 \\  
LMC--SMP 14  & 2001 Dec 09 & O6EL54010 & MIRVIS &  300 & 2 \\*  
        &             & O6EL54020 &  G750M & 1500 & 2 \\*  
        &             & O6EL54030 &  G430M &  600 & 2 \\  
LMC--SMP 29  & 2002 May 31 & O6EL5A010 & MIRVIS & 120 & 2 \\*
        &             & O6EL5A020 &  G750M & 150 & 2 \\*  
        &             & O6EL5A030 &  G430M & 210 & 2 \\  
LMC--SMP 37  & 2001 Sep 03 & O6EL5C010 & MIRVIS & 120 & 2 \\*  
        &             & O6EL5C020 &  G750M & 180 & 2 \\*  
        &             & O6EL5C030 &  G430M & 260 & 2 \\  
LMC--SMP 39  & 2002 Apr 19 & O6EL5D010 & MIRVIS & 120 & 2 \\*  
        &             & O6EL5D020 &  G750M & 480 & 2 \\*
        &             & O6EL5D030 &  G430M & 200 & 2 \\  
LMC--SMP 43  & 2001 Dec 08 & O6EL5E010 & MIRVIS & 120 & 2 \\*
        &             & O6EL5E020 &  G750M & 300 & 2 \\*  
        &             & O6EL5E030 &  G430M & 450 & 2 \\  
LMC--SMP 45  & 2002 Apr 27 & O6EL5G010 & MIRVIS & 120 & 2 \\*  
        &             & O6EL5G020 &  G750M & 980 & 2 \\*  
        &             & O6EL5G030 &  G430M & 320 & 2 \\
LMC--SMP~47  & 2002 Apr 25 & O6EL0G010 & MIRVIS & 120 & 2 \\*
        &             & O6EL0G020 &  G750M &  70 & 2 \\*
        &             & O6EL0G030 &  G750M & 330 & 2 \\*  
        &             & O6EL0G040 &  G430M &  90 & 2 \\*  
        &             & O6EL0GUJQ & MIRVIS &  15 & 2 \\  
LMC--SMP 48  & 2001 Dec 11 & O6CN15010 & MIRVIS & 120 & 2 \\*  
        &             & O6CN15020 &  G750M &  68 & 2 \\*  
        &             & O6CN15030 &  G430M & 132 & 2 \\*
        & 2002 Apr 26 & O6EL0I010 & MIRVIS & 120 & 2 \\* 
        &             & O6EL0I020 &  G750M &  30 & 2 \\*  
        &             & O6EL0I030 &  G750M & 150 & 2 \\*  
        &             & O6EL0I040 &  G750M & 600 & 2 \\*  
        &             & O6EL0I050 &  G430M &  48 & 2 \\*  
        &             & O6EL0IL4Q & MIRVIS &  15 & 1 \\  
LMC--SMP 49  & 2001 Aug 13 & O6EL5H010 & MIRVIS & 120 & 2 \\*  
        &             & O6EL5H020 &  G750M & 700 & 2 \\*  
        &             & O6EL5H030 &  G430M & 200 & 2 \\  
LMC--SMP 51  & 2001 Aug 13 & O6EL5I010 & MIRVIS & 120 & 2 \\* 
        &             & O6EL5I020 &  G750M & 300 & 2 \\*  
        &             & O6EL5I030 &  G430M & 440 & 2 \\ 
LMC--SMP 52  & 2002 Apr 11 & O6EL0M010 & MIRVIS & 120 & 2 \\* 
        &             & O6EL0M020 &  G750M &  36 & 2 \\*  
        &             & O6EL0M030 &  G750M & 180 & 2 \\*  
        &             & O6EL0M040 &  G430M &  60 & 2 \\  
LMC--SMP 57  & 2002 Jun 25 & O6EL5K010 & MIRVIS & 300 & 2 \\*  
        &             & O6EL5K020 &  G750M & 800 & 2 \\*  
        &             & O6EL5K030 &  G430M & 460 & 2 \\  
LMC--SMP 61  & 2002 Apr 27 & O6EL5M010 & MIRVIS & 120 & 2 \\*
        &             & O6EL5M020 &  G750M &  80 & 2 \\*  
        &             & O6EL5M030 &  G750M & 360 & 2 \\*  
        &             & O6EL5M040 &  G430M & 150 & 2 \\*  
        &             & O6EL5MWNQ & MIRVIS &  15 & 1 \\  
LMC--SMP 62  & 2001 Aug 31 & O6EL5N010 & MIRVIS & 120 & 2 \\*  
        &             & O6EL5N020 &  G750M &  60 & 2 \\*  
        &             & O6EL5N030 &  G750M & 300 & 2 \\*  
        &             & O6EL5N040 &  G430M &  70 & 2 \\*  
        &             & O6EL5NUAQ & MIRVIS &  15 & 1 \\  
LMC--SMP~63  & 2002 May 02 & O6EL0O010 & MIRVIS & 120 & 2 \\*
        &             & O6EL0O020 &  G750M &  34 & 2 \\*  
        &             & O6EL0O030 &  G750M & 180 & 2 \\*  
        &             & O6EL0O040 &  G430M &  60 & 2 \\*  
        &             & O6EL0OE0Q & MIRVIS &  15 & 2 \\  
LMC--SMP 64  & 2002 May 31 & O6EL5O010 & MIRVIS & 120 & 2 \\* 
        &             & O6EL5O020 &  G750M & 120 & 2 \\* 
        &             & O6EL5O030 &  G750M & 480 & 2 \\*  
        &             & O6EL5O040 &  G430M & 300 & 2 \\
LMC--SMP 67  & 2001 Nov 27 & O6EL5P010 & MIRVIS & 120 & 2 \\*  
        &             & O6EL5P020 &  G750M & 180 & 2 \\*
        &             & O6EL5P030 &  G430M & 250 & 2 \\ 
LMC--SMP 68  & 2002 Jan 28 & O6EL5Q010 & MIRVIS & 120 & 2 \\*  
        &             & O6EL5Q020 &  G750M & 600 & 2 \\*  
        &             & O6EL5Q030 &  G430M & 300 & 2 \\  
LMC--SMP 69  & 2001 Oct 16 & O6EL5R010 & MIRVIS & 120 & 2 \\*  
        &             & O6EL5R020 &  G750M & 450 & 2 \\* 
        &             & O6EL5R030 &  G430M & 360 & 2 \\
LMC--SMP 73  & 2001 Oct 23 & O6EL5U010 & MIRVIS & 120 & 2 \\*  
        &             & O6EL5U020 &  G750M &  80 & 2 \\*  
        &             & O6EL5U030 &  G750M & 360 & 2 \\*  
        &             & O6EL5U040 &  G430M & 150 & 2 \\* 
        &             & O6EL5UIYQ & MIRVIS &  15 & 1 \\
LMC--SMP 74  & 2002 Apr 28 & O6EL5V010 & MIRVIS & 120 & 2 \\* 
        &             & O6EL5V020 &  G750M &  90 & 2 \\*  
        &             & O6EL5V030 &  G430M & 150 & 2 \\*  
        &             & O6EL5VOKQ & MIRVIS &  15 & 1 \\*  
        &             & O6EL5VONQ &  G750M & 360 & 1 \\ 
LMC--SMP 75  & 2001 Aug 29 & O6EL5W010 & MIRVIS & 120 & 2 \\*  
        &             & O6EL5W020 &  G750M & 300 & 2 \\*  
        &             & O6EL5W030 &  G430M & 150 & 2 \\  
LMC--SMP 82  & 2002 May 14 & O6EL5X010 & MIRVIS & 120 & 2 \\*  
        &             & O6EL5X020 &  G750M & 900 & 2 \\*  
        &             & O6EL5X030 &  G430M & 440 & 2 \\ 
LMC--SMP 83  & 2002 Feb 28 & O6EL5Y010 & MIRVIS & 120 & 2 \\* 
        &             & O6EL5Y020 &  G750M &  50 & 2 \\*  
        &             & O6EL5Y030 &  G750M & 240 & 2 \\*  
        &             & O6EL5Y040 &  G430M &  90 & 2 \\*  
        &             & O6EL5YP1Q & MIRVIS &  15 & 1 \\
LMC--SMP 84  & 2002 Jan 05 & O6EL5Z010 & MIRVIS & 120 & 2 \\*  
        &             & O6EL5Z020 &  G750M & 150 & 2 \\*  
        &             & O6EL5Z030 &  G430M & 270 & 2 \\
LMC--SMP 88  & 2001 Aug 30 & O6EL62010 & MIRVIS & 120 & 2 \\*  
        &             & O6EL62020 &  G750M & 320 & 2 \\*
        &             & O6EL62030 &  G430M & 540 & 2 \\  
LMC--SMP 89  & 2001 Aug 28 & O6EL63010 & MIRVIS & 120 & 2 \\* 
        &             & O6EL63020 &  G750M &  45 & 2 \\*  
        &             & O6EL63030 &  G750M & 180 & 2 \\*
        &             & O6EL63040 &  G430M &  72 & 2 \\* 
        &             & O6EL63F3Q & MIRVIS &  15 & 1 \\
LMC--SMP 91  & 2002 Jun 30 & O6EL65010 & MIRVIS &  120 & 2 \\*  
        &             & O6EL65020 &  G750M & 1100 & 2 \\*   
        &             & O6EL65030 &  G430M &  600 & 2 \\ 
LMC--SMP 92  & 2002 Jun 10 & O6EL66010 & MIRVIS & 120 & 2 \\* 
        &             & O6EL66020 &  G750M &  74 & 2 \\* 
        &             & O6EL66030 &  G750M & 360 & 2 \\* 
        &             & O6EL66040 &  G430M & 160 & 2 \\*  
        &             & O6EL66MAQ & MIRVIS &  15 & 1 \\ 
LMC--SMP 98  & 2002 Jan 29 & O6EL67010 & MIRVIS & 120 & 2 \\*  
        &             & O6EL67020 &  G750M & 120 & 2 \\*
        &             & O6EL67030 &  G750M & 600 & 2 \\* 
        &             & O6EL67040 &  G430M & 110 & 2 \\
LMC--SMP 101 & 2001 Aug 17 & O6EL68010 & MIRVIS & 120 & 2 \\* 
        &             & O6EL68020 &  G750M & 200 & 2 \\*
        &             & O6EL68030 &  G430M & 270 & 2 \\  
\cutinhead{SMC Nebulae}
SMC--SMP 2   & 2004 Jul 22 & O8Y040010 & MIRVIS &  120 & 2 \\*
        &             & O8Y040020 &  G750M &  150 & 2 \\*  
        &             & O8Y040030 &  G750M &  600 & 2 \\*  
        &             & O8Y040040 &  G430M &  120 & 2 \\ 
SMC--SMP 3   & 2004 Jul 17 & O8Y041010 & MIRVIS &  120 & 2 \\*
        &             & O8Y041020 &  G750M & 1200 & 2 \\*
        &             & O8Y041030 &  G430M &  900 & 2 \\ 
SMC--SMP 15  & 2004 Jul 13 & O8Y046010 & MIRVIS &  120 & 2 \\*
        &             & O8Y046020 &  G750M &   60 & 2 \\*  
        &             & O8Y046030 &  G750M &  240 & 2 \\*  
        &             & O8Y046040 &  G430M &  210 & 2 \\ 
SMC--SMP 16  & 2004 Jul 22 & O8Y047010 & MIRVIS &  120 & 2 \\*
        &             & O8Y047020 &  G750M &  120 & 2 \\*  
        &             & O8Y047030 &  G750M &  480 & 2 \\*  
        &             & O8Y047040 &  G430M &  360 & 2 \\ 
SMC--SMP 28  & 2004 Jul 14 & O8Y052010 & MIRVIS &  120 & 2 \\*
        &             & O8Y052020 &  G430M & 1200 & 2 \\ 
\enddata
\end{deluxetable} 

%
%
\begin{deluxetable}{lrllllrc}
\tabletypesize{\scriptsize}
\tablecolumns{6}
\tablewidth{0pt}
\tablecaption {Log of HST Images from Other Programs \label{Serendip}}
\tablehead {
\colhead {} & \colhead {GO} & \colhead {} & \colhead {} & 
\colhead {Instrument/} & \colhead {} & \colhead {T$_{Exp}$} & \colhead {} \\
\colhead {Nebula} & \colhead {Program} & \colhead {Date} & \colhead {Dataset} & 
\colhead {Aperture} & \colhead {Filter} & \colhead {(s)} & 
\colhead {N$_{Exp}$} 
}
\startdata  
\cutinhead{LMC Nebulae}
LMC--MA~17  &  7303 & 1999 Jan 6  & O57N01GLQ & STIS/50CCD & MIRVIS &  200 & 1 \\
LMC--MG~60  &  9584 & 2003 Feb 22 & J8G9SDD6Q & ACS/WFC    & F775W &  750 & 1 \\*
            &       &             & J8G9SDDBQ &            & F775W &  604 & 1 \\
LMC--Mo~17  &  9891 & 2003 Oct 8  & J8NE55Z9Q & ACS/WFC1   & F555W &  300 & 1 \\*
            &       &             & J8NE55ZEQ &            & F814W &  200 & 1 \\
LMC--Mo~36  &  9285 & 2001 Dec 11 & O6J3LGT1Q & STIS/50CCD & MIRVIS & 200 & 1 \\*
            &       &             & O6J3LGT2Q &            & MIRVIS & 200 & 1 \\
LMC--RP~265 &  8059 & 2001 Aug 17 & U4WOE702B & WFPC2/WFALL & F606W & 240 & 2 \\
\cutinhead{SMC Nebulae}
SMC--JD~12  & 10126 & 2004 Oct 30 & J90604KSQ & ACS/WFC1 & F435W &  440 & 1 \\*
            &       &             & J90604KTQ &          & F555W &  440 & 1 \\* 
            &       &             & J90604KWQ &          & F555W &  120 & 1 \\* 
            &       &             & J90604KXQ &          & F814W &  440 & 1 \\*
            &       &             & J90604L0Q &          & F814W &  120 & 1 \\ 
\enddata
\end{deluxetable} 

%
%
\begin{deluxetable}{lccc}
\tabletypesize{\scriptsize}
\tablewidth{0pt}
\tablecolumns{4}
\tablecaption {Spectrum Extraction Parameters \label{SpExtr}}
\tablehead {
\colhead {} & \multicolumn{2}{c}{Center} & \colhead {} \\
\cline{2-3} \\
\colhead {} & \colhead {G430M} & \colhead {G750M} & \colhead {Width} \\
\colhead {Nebula} & \colhead {(pixel)} & \colhead {(pixel)} & \colhead {(pixel)}
}
\startdata  
\cutinhead{LMC Nebulae}
LMC--J 5     & 517 & 514 & 43 \\
LMC--J 33    & 496 & 492 & 33 \\
LMC--MG 4    & 470 & 466 & 67 \\
LMC--MG 14   & 516 & 512 & 37 \\
LMC--MG 16   & 520 & 518 & 55 \\
LMC--MG 29   & 491 & 487 & 49 \\
LMC--MG 40   & 537 & 533 & 29 \\
LMC--MG 45   & 562 & 558 & 9 \\
LMC--MG 51   & 510 & 505 & 29 \\
LMC--MG 70   & 509 & 504 & 25 \\
LMC--Mo 7    & 495 & 492 & 21 \\
LMC--Mo 21   & 523 & 520 & 83 \\
LMC--Mo 33   & 501 & 496 & 83 \\
LMC--Mo 36   & 494 & 489 & 19 \\
LMC--Mo 47   & 515 & 513 & 69 \\
LMC--Sa 104a & 534 & 530 & 13 \\
LMC--Sa 107  & 529 & 525 & 45 \\
LMC--Sa 117  & 515 & 511 & 31 \\
LMC--Sa 121  & 548 & 543 & 41 \\
LMC--SMP 3   & 493 & 489 & 11 \\
LMC--SMP 5   & 502 & 498 & 15 \\
LMC--SMP 6   & 535 & 531 & 15 \\  
LMC--SMP 11  & 496 & 494 & 19 \\  
LMC--SMP 14  & 507 & 503 & 21 \\  
LMC--SMP 29  & 507 & 503 & 21 \\  
LMC--SMP 37  & 483 & 479 & 15 \\  
LMC--SMP 39  & 518 & 513 & 15 \\
LMC--SMP 43  & 557 & 553 & 25 \\  
LMC--SMP 45  & 522 & 518 & 51 \\  
LMC--SMP~47  & 565 & 561 & 21 \\
LMC--SMP 48  & 507 & 502 & 15 \\
LMC--SMP 49  & 535 & 530 & 31 \\  
LMC--SMP 51  & 543 & 539 & 11 \\  
LMC--SMP 52  & 527 & 522 & 19 \\
LMC--SMP 57  & 510 & 507 & 23 \\
LMC--SMP 61  & 547 & 542 & 21 \\  
LMC--SMP 62  & 525 & 520 & 27 \\  
LMC--SMP~63  & 531 & 528 & 19 \\
LMC--SMP 64  & 510 & 505 & 15 \\  
LMC--SMP 67  & 472 & 468 & 27 \\
LMC--SMP 68  & 532 & 538 & 31 \\   
LMC--SMP 69  & 520 & 516 & 61 \\ 
LMC--SMP 73  & 496 & 492 & 23 \\  
LMC--SMP 74  & 489 & 484 & 21 \\  
LMC--SMP 75  & 513 & 509 & 17 \\  
LMC--SMP 82  & 502 & 499 & 19 \\   
LMC--SMP 83  & 521 & 517 & 75 \\  
LMC--SMP 84  & 480 & 476 & 21 \\  
LMC--SMP 88  & 522 & 518 & 17 \\
LMC--SMP 89  & 541 & 536 & 21 \\
LMC--SMP 91  & 546 & 542 & 115 \\ 
LMC--SMP 92  & 527 & 523 & 23 \\
LMC--SMP 98  & 455 & 451 & 13 \\
LMC--SMP 101 & 540 & 536 & 47 \\
\cutinhead{SMC Nebulae}
SMC--SMP 2  & 514 & 509 & 20 \\
SMC--SMP 3  & 518 & 514 & 20 \\
SMC--SMP 15 & 495 & 492 & 14 \\
SMC--SMP 16 & 546 & 542 & 15 \\
SMC--SMP 28 & 462 & \nodata & 13 \\ 
\enddata
\end{deluxetable} 

%
%
\begin{deluxetable}{llrrrrrrrrrrrr}
\tabletypesize{\scriptsize}
\rotate
\tablecolumns{14}
\tablewidth{0pc}
\tablecaption {Relative Emission Line Intensities of Magellanic Cloud Nebulae\label{Flux}}
\tablehead {
\colhead {} & \colhead {log F(H$\beta$)} & \colhead {} & \colhead {[O~III]} 
& \colhead {[O~III]} & \colhead {[O~I]} & \colhead {[S~III]} & 
\colhead {[O~I]} & \colhead {[N II]} & \colhead {H$\alpha$} & 
\colhead {[N~II]} & \colhead {He~I} & \colhead {[S~II]} & \colhead {[S~II]} \\
\colhead {Nebula} & \colhead {($\lambda$4861)} & \colhead {$c$} & \colhead {($\lambda$4959)} & 
\colhead {($\lambda$5007)} & \colhead {($\lambda$6300)} & \colhead {($\lambda$6312)} & 
\colhead {($\lambda$6363)} & \colhead {($\lambda$6548)} & \colhead {($\lambda$6563)} & 
\colhead {($\lambda$6583)} & \colhead {($\lambda$6678)} & \colhead {($\lambda$6717)} & 
\colhead {($\lambda$6731)} \\
\colhead {(1)} & \colhead {(2)} & \colhead {(3)} & \colhead {(4)} & 
\colhead {(5)} & \colhead {(6)} & \colhead {(7)} & \colhead {(8)} & 
\colhead {(9)} & \colhead {(10)} & \colhead {(11)} & \colhead {(12)} & 
\colhead {(13)} & \colhead {(14)}  
} 
\startdata  
\cutinhead{LMC Nebulae}
LMC-J~5    &$-13.46\tablenotemark{a}$ &\nodata &166.2  &488.2   &\nodata &\nodata &\nodata &\multicolumn{3}{c}{1298.\tablenotemark{a,b}} &\nodata &37.0\tablenotemark{c} &33.2\tablenotemark{c} \\
LMC-J 33   &$-13.79$ &0.00   &312.9   &922.2   &\nodata &\nodata &\nodata &\nodata &281.7   &\nodata &\nodata &\nodata &\nodata \\
LMC-MG~4   &$-14.13\tablenotemark{d}$ &\nodata &260.0 &797.9   &\nodata &\nodata &\nodata &\multicolumn{3}{c}{1975.\tablenotemark{b,e}} &\nodata &\multicolumn{2}{c}{237.9\tablenotemark{f}} \\
LMC-MG 14  &$-13.58$ &0.0    &358.3   &1053.4  &\nodata &\nodata &\nodata &\nodata &280.3   &\nodata &0.0     &0.0     &0.0 \\
LMC-MG 16  &$-13.95$ &\nodata &234.0  &709.3   &\nodata &\nodata &\nodata &\multicolumn{3}{c}{1545.\tablenotemark{b}} &\nodata &\multicolumn{2}{c}{124.2\tablenotemark{f}} \\
LMC-MG 29  &$-13.54$ &0.15:  &359.7   &1045.2  &11.2    &\nodata &3.0     &44.9    &320.8\tablenotemark{e}   &134.3   &\nodata &\multicolumn{2}{c}{20.6\tablenotemark{f}} \\
LMC-MG 40  &$-13.68$ &0.17   &247.4   &725.6   &\nodata &\nodata &\nodata &\nodata &326.9   &\nodata &\nodata &\nodata &\nodata \\
LMC-MG 45  &$-13.15$ &0.78   &477.3   &1406.0  &12.8    &3.5     &4.6     &12.1    &536.4   &38.8    &7.6     &1.4     &2.6 \\
LMC-MG 51  &$-14.03$ &0.00   &250.5   &763.8   &\nodata &\nodata &\nodata &\nodata &279.9   &\nodata &\nodata &\nodata &\nodata \\
LMC-MG 70  &$-13.79$ &0.19   &323.2   &958.5   &20.3    &3.4     &7.4     &99.3	   &331.3   &303.2   &6.5     &15.7    &19.1 \\
LMC-Mo 07  &$-14.20$ &0.01   &193.2   &569.4   &\nodata &\nodata &\nodata &\nodata &286.3   &\nodata &\nodata &\nodata &\nodata \\
LMC-Mo 21  &$-14.37$ &\nodata &139.5  &434.0   &\nodata &\nodata &\nodata &\multicolumn{3}{c}{2828.\tablenotemark{b,e}} &\nodata &\nodata &\nodata \\
LMC-Mo 33  &$-13.79$ &\nodata &326.1  &991.3   &\nodata &\nodata &\nodata &\multicolumn{3}{c}{447.2\tablenotemark{b}} &\nodata &\multicolumn{2}{c}{42.0\tablenotemark{f}} \\
LMC-Mo 36  &$-14.26$ &0.29   &272.7   &779.5   &\nodata &\nodata &18.9    &139.4   &360.4   &416.5   &\nodata &56.9    &55.7 \\
LMC-Mo 47  &$-14.0\tablenotemark{d}$ &\nodata &127.4   &378.9   &\nodata &\nodata &\nodata &\multicolumn{3}{c}{221.1\tablenotemark{b}} &\nodata &\nodata &\nodata \\
LMC-Sa 104a &$-12.66$ &0.07  &61.0    &184.6   &3.1     &1.9     &1.1     &1.3	   &303.9   &3.6     &1.3     &0.35    &0.54 \\
LMC-Sa 107 &$-13.91$ &0.51   &282.9   &1057.4  &\nodata &\nodata &\nodata &432.1   &430.8   &1295.5  &6.4:    &\multicolumn{2}{c}{167.3\tablenotemark{f}} \\
LMC-Sa 117 &$-13.64$ &0.21   &217.7   &643.5   &\nodata &\nodata &\nodata &\nodata &338.8   &\nodata &\nodata &\nodata &\nodata \\
LMC-Sa 121 &$-13.83$ &0.00   &247.1   &718.9   &31.5    &\nodata &\nodata &78.4    &284.3   &234.6   &\nodata &36.9\tablenotemark{c}    &34.0\tablenotemark{c} \\
LMC-SMP~3  &$-12.48$ &0.01   &134.3   &395.3   &1.4     &0.6     &0.6     &2.7     &286.3   &8.9     &4.0     &0.2     &0.6 \\
LMC-SMP~5  &$-12.88$ &0.03   &96.0    &298.6   &\nodata &\nodata &\nodata &8.4     &291.7   &25.3    &3.8     &1.7     &2.6 \\
LMC-SMP~6  &$-12.89$ &0.69   &581.4   &1821.7  &13.6    &\nodata &\nodata &18.9	   &498.4   &55.0    &\nodata &\nodata &8.1 \\
LMC-SMP 11 &$-13.94$ &0.31   &16.0    &60.7    &163.6   &\nodata &53.6    &99.7	   &366.1   &315.7   &\nodata &34.6    &72.3 \\
LMC-SMP 14 &$-13.72$ &\nodata &255.0  &773.8   &31.1    &\nodata &\nodata &\multicolumn{3}{c}{1675.\tablenotemark{b,e}} &2.1:    &\multicolumn{2}{c}{129.1\tablenotemark{f}} \\
LMC-SMP 29 &$-12.74$ &0.21   &336.5   &991.7   &\nodata &5.4     &9.4     &142.7   &337.7   &437.8   &3.7     &14.7    &26.2 \\
LMC-SMP 37 &$-12.92$ &0.20   &393.6   &1169.8  &11.5    &3.2     &3.3     &74.6	   &334.5   &237.0   &2.5     &7.2     &12.0 \\
LMC-SMP 39 &$-13.27$ &0.31   &386.1   &1164.3  &20.5    &2.7     &7.1     &49.0	   &367.0   &149.5   &3.9     &16.2    &27.0 \\
LMC-SMP 43 &$-13.12$ &0.15   &374.2   &1116.5  &\nodata &\nodata &\nodata &\nodata &320.5   &\nodata &\nodata &\nodata &\nodata \\
LMC-SMP 45 &$-13.13$ &0.40   &406.2   &1219.8  &\nodata &\nodata &4.6     &36.1	   &393.5   &105.2   &4.5     &\multicolumn{2}{c}{44.0\tablenotemark{f}} \\
LMC-SMP 47 &$-12.54$ &0.23   &340.8   &1027    &12.5    &4.4     &4.0     &99.9	   &343.8   &307.3   &3.7     &5.2     &10.7 \\
LMC-SMP 48 &$-12.48$ &0.27   &227.2   &679.8   &3.6     &1.3     &1.1     &9.1	   &352.9   &27.3    &4.8     &1.1     &2.4 \\
LMC-SMP 49 &$-13.21$ &0.11:  &345.3   &1031.3  &\multicolumn{2}{c}{15.6\tablenotemark{g}}   &4.1     &50.0:\tablenotemark{h}   &310.3\tablenotemark{h}   &141.1\tablenotemark{h}   &2.8     &32.9    &34.4 \\
LMC-SMP 51 &$-13.06$ &0.86   &346.6   &1044.1  &4.7     &2.3     &1.6     &6.2	   &568.5   &21.5    &7.7     &0.7     &1.3 \\
LMC-SMP 52 &$-12.63$ &0.24   &468.5   &1440.   &4.0     &2.5     &1.2     &8.0	   &345.8   &24.4    &4.5     &3.8     &6.9 \\
LMC-SMP 57 &$-13.61$ &0.19   &337.2   &1034.7\tablenotemark{i} &\nodata &\nodata &\nodata &\nodata &333.3\tablenotemark{i} &\nodata &3.6:    &\nodata &\nodata \\
LMC-SMP 61 &$-12.48$ &0.22   &255.1   &762.7   &\nodata &1.4     &1.6     &10.9	   &340.9   &34.5    &4.8     &2.2\tablenotemark{c}     &4.3\tablenotemark{c} \\
LMC-SMP 62 &$-12.30$ &0.07   &343.6   &1032.9  &6.2     &3.5     &2.1     &13.2    &302.7   &38.7    &3.2     &3.3     &5.7 \\
LMC-SMP 63 &$-12.46$ &0.14   &307.3   &949.3   &2.9     &2.1     &1.2     &6.2	   &319.3   &19.0    &4.5     &1.3     &2.4 \\
LMC-SMP 64 &$-12.78$ &1.12   &7.5     &22.4    &6.8     &5.9     &2.5     &12.4    &704.9   &44.8    &2.9     &0.8     &0.8 \\
LMC-SMP 67 &$-12.78$ &0.14   &101.6   &304.6   &1.7     &1.2:    &\nodata &87.0    &317.9   &265.7   &4.4     &2.7     &4.3 \\
LMC-SMP 68 &$-13.17$ &0.00   &218.8   &628.7   &\nodata &\nodata &\nodata &\nodata &281.3   &\nodata &\nodata &\nodata &\nodata \\
LMC-SMP 69 &$-13.43$ &\nodata &285.9  &872.1   &42.3:   &\nodata &8.9:    &\multicolumn{3}{c}{1509.5\tablenotemark{b,e}} &\nodata &\multicolumn{2}{c}{104.9\tablenotemark{f}} \\
LMC-SMP 73 &$-12.55$ &0.17   &486.5   &1458.3  &11.2    &2.6     &4.3     &19.0	   &327.1   &56.6    &4.1:\tablenotemark{j}     &3.3:\tablenotemark{j}     &7.2:\tablenotemark{j} \\
LMC-SMP 74 &$-12.64$ &0.09   &385.6   &1146.3  &7.4     &2.4:    &2.5:    &9.3	   &306.2   &36.2    &4.3     &3.3     &7.4 \\
LMC-SMP 75 &$-12.53$ &0.26   &339.2   &979.1   &4.8     &1.5     &1.7     &8.6     &350.7   &26.4    &5.3     &1.0     &2.2 \\
LMC-SMP 82 &$-13.62$ &0.47   &375.4   &1156.8  &14.5    &8.5     &5.3     &125.3   &417.4   &372.2   &9.3     &15.0    &23.2 \\
LMC-SMP 83 &$-12.67$ &\nodata &282.7  &800.2   &\multicolumn{2}{c}{8.5\tablenotemark{g}} &\nodata &\multicolumn{3}{c}{507.5\tablenotemark{b}} &\nodata  &\multicolumn{2}{c}{22.4\tablenotemark{f}} \\
LMC-SMP 84 &$-12.68$ &0.08   &229.9   &682.7   &\nodata &1.7:    &\nodata &2.7:    &303.6   &6.3     &3.4     &\nodata &\nodata \\
LMC-SMP 88 &$-13.44$ &0.58   &113.5   &333.0   &9.8     &3.5     &3.7     &50.0    &453.8   &155.2   &5.8:    &7.9     &13.1 \\
LMC-SMP 89 &$-12.61$ &0.31   &431.7   &1309.9  &9.7     &2.4     &3.2     &13.1    &366.6   &40.3    &4.4     &3.3     &6.3 \\
LMC-SMP 91 &$-13.59$ &\nodata &282.7  &852.9   &\multicolumn{2}{c}{48.6\tablenotemark{g}}   &14.4    &\multicolumn{3}{c}{1902.9\tablenotemark{b,e}} &\nodata &\multicolumn{2}{c}{145.5:\tablenotemark{f}} \\
LMC-SMP 92 &$-12.58$ &0.16   &466.8   &1386.6  &\nodata &2.7     &3.8     &20.4	   &325.4   &64.1    &3.5     &5.3     &9.4 \\
LMC-SMP 98 &$-12.54$ &0.28   &478.5   &1395.5  &10.3:   &4.3     &3.6     &13.3	   &356.7   &41.0    &4.2     &2.0     &4.7 \\
LMC-SMP 101 &$-12.93$ &0.08: &387.4   &1138.1  &\nodata &\nodata &\nodata &4.1\tablenotemark{h} &303.5\tablenotemark{h} &11.9\tablenotemark{h} &\nodata &\nodata &\nodata \\
\cutinhead{SMC Nebulae}
SMC-SMP 02 &$-12.71$ &0.00   &303.6   &909.0   & 2.8    & 1.4    & 0.9    & 3.5    &283.0   &10.0    & 3.0    &1.4     & 2.4 \\
SMC-SMP 03 &$-13.13$ &0.00   &272.5   &816.6   &\nodata &$<1.0$  &$<1.0$  & 1.1    &284.0   & 3.3    & 3.8    &$<1.0$  &$<1.0$ \\
SMC-SMP 15 &$-12.45$ &0.01   &208.2   &584.4   & 1.9    & 0.8    & 0.7    & 4.4    &286.8   &13.5    & 4.5    & 0.4    & 1.0 \\
SMC-SMP 16 &$-12.74$ &0.03   & 57.2   &165.5   & 1.4    & 0.7    & 0.5    &11.6    &291.9   &35.5    & 3.2    & 1.1    & 2.3 \\
SMC-SMP 28 &$-13.18$ &\nodata &93.6   &278.7   &\nodata &\nodata &\nodata &\nodata &\nodata &\nodata &\nodata &\nodata &\nodata \\
\enddata
\tablenotetext{a}{Bulk of H$^+$ emission is probably from CS (see text).}
\tablenotetext{b}{Intensity of the blend of H$\alpha$ and [\ion{N}{2}] $\lambda$6548,6583.}
\tablenotetext{c}{Partial blending of doublet [\ion{S}{2}] $\lambda$6716,6731.}
\tablenotetext{d}{Extremely weak surface brightness in this line renders flux measurement extremely uncertain.}
\tablenotetext{e}{Intensity of [\ion{N}{2}] $\lambda\lambda$6548,6583 dominates emission from H$\alpha$ (see text).}
\tablenotetext{f}{Intensity of the blended doublet [\ion{S}{2}] $\lambda$6716,6731.}
\tablenotetext{g}{Intensity of the blend of [\ion{O}{1}] $\lambda$6300 and [\ion{S}{3}] $\lambda$6312.}
\tablenotetext{h}{Partial blending of [\ion{N}{2}] $\lambda$6548 and H$\alpha$.}
\tablenotetext{i}{Approximately 7\% of the emission is contained in the plume (see text).}
\tablenotetext{j}{Greater uncertainty due to complicated stellar continuum.}
\end{deluxetable} 

%
%
\begin{deluxetable}{lccccccl}
\tabletypesize{\scriptsize}
\tablecolumns{8}
\tablewidth{0pc}
\tablecaption {Coordinates, Dimensions, and Morphologies of Magellanic Cloud  Nebulae\label{Morph}}
\tablehead {
\colhead {} & \colhead {R.A.} & \colhead {Decl.} & \colhead {R$_{phot}$} & 
\colhead {Diameter} & \colhead {Morph.} & \colhead {} & \colhead {} \\
\colhead {Nebula} & \colhead {J(2000)} & \colhead {(J2000)} & 
\colhead {(arcsec)} & \colhead {(arcsec)} & \colhead {Class} & 
\colhead {Figure} & \colhead {Notes} \\
\colhead {(1)} & \colhead {(2)} & \colhead {(3)} & \colhead {(4)} & 
\colhead {(5)} & \colhead {(6)} & \colhead {(7)} & \colhead {(8)} 
}
\startdata 
\cutinhead{LMC Nebulae}
LMC--J~5     & 5 11 48.05 & $-69$ 23 42.2 & 0.67 & 0.97 x 1.48 & E     & \ref{Neb_1} & Stellar emission, P-Cyg? \\
LMC--J~33    & 5 21 18.11 & $-69$ 43 01.9 & 0.69 & 1.53 x 1.79 & E     & \ref{Neb_1} & Attached shell   \\ 
LMC--MA~17   & 5 55 13.78 & $-65$ 28 23.5 & \nodata & 0.88 x 0.67 & Q  & \ref{Neb_16} & Jets \\ 
LMC--MG~4    & 4 52 44.83 & $-70$ 17 50.6 & 1.19 & 4.3 x 3.3   & E?    & \ref{Neb_2} & Faint; could be B?\\ 
LMC--MG~14   & 5 04 27.67 & $-68$ 58 12.3 & 0.56 &     1.58    & R(bc) & \ref{Neb_1} & Attached shell \\ 
LMC--MG 16   & 5 06 05.22 & $-64$ 48 48.9 & 0.62 & 1.28 x 1.63 & B     & \ref{Neb_1} & \\ 
LMC--MG 29   & 5 13 42.48 & $-68$ 15 17.9 & 1.21 & 1.48 x 2.30 & E     & \ref{Neb_1} & \\ 
LMC--MG 40   & 5 22 35.36 & $-68$ 24 26.5 & 0.70 & 0.38 x 0.33 & E(bc) & \ref{Neb_3} & Attached shell \\ 
LMC--MG 45   & 5 26 06.45 & $-63$ 24 04.4 & 0.20 & 0.31 x 0.23 & E     & \ref{Neb_3} &   \\ 
LMC--MG 51   & 5 28 34.47 & $-70$ 33 01.9 & 0.62 & 1.22 x 1.43 & E     & \ref{Neb_3} & Attached shell \\ 
LMC--MG 60   & 5 33 30.95 & $-69$ 08 13.6 & \nodata & 1.20 x 1.15 & R  & \ref{Neb_16} & \\ 
LMC--MG 70   & 5 38 12.37 & $-75$ 00 21.6 & 0.31 & 0.48 x 0.67 & E(bc) & \ref{Neb_3} &    \\ 
LMC--Mo~7    & 4 49 19.86 & $-64$ 22 36.8 & 0.43 & 0.72 x 0.93 & E     & \ref{Neb_3} &   \\ 
LMC--Mo 17   & 5 07 35.06 & $-67$ 29 02.1 & \nodata & 2.07 x 1.54 & E  & \ref{Neb_16} & Attached shell; inner core: 0\farcs64x0\farcs58 \\ 
LMC--Mo 21   & 5 19 04.11 & $-64$ 44 39.3 & 1.06 &  3.1 x 2.9  & B     & \ref{Neb_2} & \\ 
LMC--Mo 33   & 5 32 09.75 & $-70$ 24 43.7 & 1.93 & 2.12 x 1.58 & E(bc?) & \ref{Neb_4} & \\ 
LMC--Mo 36   & 5 38 53.81 & $-69$ 57 56.0 & 0.34 & 1.14 x 0.97 & E     & \ref{Neb_5} & \\ 
LMC--Mo 47   & 6 13 03.35 & $-65$ 55 09.4 & 1.28 &     3.47    & R     & \ref{Neb_4} & v. faint \\ 
LMC--RP~265  & 5 37 00.7  & $-69$ 21 29. & \nodata & 4.2 x 3.4 & B?    & \ref{Neb_16} & v. faint \\ 
LMC--Sa 104a & 4 25 32.15 & $-66$ 47 18.5 & \nodata & $<0.25$  & \nodata & \ref{Neb_5} & Resolved velocity structure \\
LMC--Sa 107  & 5 06 43.81 & $-69$ 15 38.4 & 0.75 & 1.70 x 1.62 & B?    & \ref{Neb_5} & Could be P? \\
LMC--Sa 117  & 5 24 56.82 & $-69$ 15 31.2 & 0.77 & 1.18 x 1.30 & P     & \ref{Neb_5} & \\ 
LMC--Sa 121  & 5 30 26.20 & $-71$ 13 49.5 & 0.58 & 1.58 x 1.65 & E     & \ref{Neb_5} & Ansae; attached shell \\ 
LMC--SMP~3   & 4 42 23.79 & $-66$ 13 01.0 & 0.15 & 0.26 x 0.23 & R?    & \ref{Neb_6} & Barely resolved \\ 
LMC--SMP~5   & 4 48 08.62 & $-67$ 26 06.5 & 0.27 & 0.46 x 0.50 & E     & \ref{Neb_6} & \\ 
LMC--SMP~6   & 4 47 38.97 & $-72$ 28 21.6 & 0.25 & 0.67 x 0.56 & E(bc) & \ref{Neb_6} & \\ 
LMC--SMP~11  & 4 51 37.69 & $-67$ 05 16.1 & 0.43 & 0.76 x 0.55 & B     & \ref{Neb_6} & Detached arc \\ 
LMC--SMP~14  & 5 00 21.09 & $-70$ 58 52.3 & 1.33 & 2.41 x 1.87 & B     & \ref{Neb_7} & \\ 
LMC--SMP~29  & 5 08 03.34 & $-68$ 40 16.8 & 0.27 & 0.51 x 0.47 & BC    & \ref{Neb_6} & Irregular, B? \\ 
LMC--SMP~37  & 5 11 03.12 & $-67$ 47 57.6 & 0.25 & 0.50 x 0.43 & E     & \ref{Neb_8} & \\ 
LMC--SMP~39  & 5 11 42.26 & $-68$ 34 59.3 & 0.23 & 0.60 x 0.55 & E     & \ref{Neb_8} & Ring? \\ 
LMC--SMP~43  & 5 17 02.40 & $-69$ 07 16.1 & 0.55 & 1.11        & R(bc) & \ref{Neb_8} & Attached shell \\ 
LMC--SMP~45  & 5 19 20.75 & $-66$ 58 08.4 & 0.72 & 1.66 x 1.62 & B?    & \ref{Neb_7} & \\ 
LMC--SMP~47  & 5 19 54.88 & $-69$ 31 04.3 & 0.25 & 0.45 x 0.32 & E     & \ref{Neb_8} & J~25; ansa to SSE \\ 
LMC--SMP~48  & 5 20 09.66 & $-69$ 53 39.2 & 0.20 & 0.40 x 0.35 & E     & \ref{Neb_8} & J~27 \\ 
LMC--SMP~49  & 5 20 09.46 & $-70$ 25 38.5 & 0.65 & 1.00        & R     & \ref{Neb_9} &   \\ 
LMC--SMP~51  & 5 20 52.56 & $-70$ 09 36.6 & \nodata & \nodata & \nodata & \ref{Neb_9} & Unresolved \\ 
LMC--SMP~52  & 5 21 23.75 & $-68$ 35 34.9 & 0.20 &     0.73    & R     & \ref{Neb_9} & J~34 \\ 
LMC--SMP~57  & 5 23 48.69 & $-69$ 12 20.9 & 0.43 & 0.93 x 0.90 & R(bc) & \ref{Neb_9} & Asymmetric core; attached shell; plume \\ 
LMC--SMP~61  & 5 24 36.37 & $-73$ 40 40.0 & 0.26 & 0.56 x 0.54 & E     & \ref{Neb_9} & \\ 
LMC--SMP~62  & 5 24 55.24 & $-71$ 32 56.4 & 0.27 & 0.59 x 0.41 & E(bc) & \ref{Neb_10} & \\ 
LMC--SMP~63  & 5 25 26.12 & $-68$ 55 55.4 & 0.26 & 0.63 x 0.57 & E     & \ref{Neb_10} & J~39 \\ 
LMC--SMP~64  & 5 27 35.69 & $-69$ 08 56.4 & \nodata & $<0.25$ & \nodata & \ref{Neb_10} & Stellar emission? \\
LMC--SMP~67  & 5 29 16.01 & $-67$ 32 48.0 & 0.33 & 0.88 x 0.61 & B     & \ref{Neb_10} & \\ 
LMC--SMP~68  & 5 29 02.83 & $-70$ 19 23.8 & 0.71 & 1.33 x 0.97 & E(s)  & \ref{Neb_10} & \\ 
LMC--SMP~69  & 5 29 23.31 & $-67$ 13 21.9 & 1.03 & 1.84 x 1.43 & B     & \ref{Neb_11} &  \\ 
LMC--SMP~73  & 5 31 22.27 & $-70$ 40 45.3 & 0.25 & 0.31 x 0.27 & E(bc) & \ref{Neb_11} & \\ 
LMC--SMP~74  & 5 33 29.51 & $-71$ 52 27.9 & 0.35 & 0.79 x 0.63 & E(bc) & \ref{Neb_11} & Might be B? \\ 
LMC--SMP~75  & 5 33 47.00 & $-68$ 36 44.0 & 0.20 &     0.33    & R     & \ref{Neb_11} & \\ 
LMC--SMP~82  & 5 35 57.47 & $-69$ 58 16.6 & 0.20 & 0.31 x 0.30 & E     & \ref{Neb_11} & \\ 
LMC--SMP~83  & 5 36 20.82 & $-67$ 18 07.0 & 1.26 & 3.98 x 3.63 & B?    & \ref{Neb_12} & Could be P? \\ 
LMC--SMP~84  & 5 36 53.17 & $-71$ 53 39.8 & 0.29 & 0.57 x 0.48 & E     & \ref{Neb_13} & \\ 
LMC--SMP~88  & 5 42 33.27 & $-70$ 29 24.2 & 0.36 & 0.61 x 0.45 & E     & \ref{Neb_13} & \\ 
LMC--SMP~89  & 5 42 36.84 & $-70$ 09 32.0 & 0.23 & 0.51 x 0.45 & E(bc?) & \ref{Neb_13} & \\ 
LMC--SMP~91  & 5 45 06.02 & $-68$ 06 49.1 & 1.12 & 1.89 x 1.40 & B     & \ref{Neb_12} & \\ 
LMC--SMP~92  & 5 47 04.63 & $-69$ 27 34.5 & 0.26 & 0.62 x 0.54 & E(bc) & \ref{Neb_13} & \\ 
LMC--SMP~98  & 6 17 35.59 & $-73$ 12 37.3 & 0.20 &     0.41    & R     & \ref{Neb_13} & \\ 
LMC--SMP~101 & 6 23 40.24 & $-69$ 10 38.9 & 0.81 & 1.03 x 0.82 & E(bc?) & \ref{Neb_14} & Ansae \\
\cutinhead{SMC Nebulae}
SMC--JD~12   & 0 51 07.31 & $-73$ 12 04.4 & \nodata & 1.15 x 1.40 & B? & \ref{Neb_16} & \\
SMC--SMP~2  & 0 32 38.81 & $-71$ 41 58.7 & 0.25 & 0.54        & R     & \ref{Neb_15} & \\
SMC--SMP~3  & 0 34 21.89 & $-73$ 13 20.4 & 0.27 & 0.59 x 0.48 & E(bc) & \ref{Neb_15} & \\ 
SMC--SMP~15  & 0 51 07.45 & $-73$ 57 37.1 & 0.17 & 0.32        & R     & \ref{Neb_15} & \\
SMC--SMP~16  & 0 51 27.08 & $-72$ 26 11.1 & 0.18 & 0.33 x 0.30 & E     & \ref{Neb_15} & \\
SMC--SMP~28  & 1 24 12.09 & $-74$ 02 31.4 & 0.21 & 0.31        & R     & \ref{Neb_15} & Ansa? \\
\enddata
\end{deluxetable} 

\clearpage

\centerline{\bf Figure Captions}

\figcaption[f4.eps]{
Images ({\it left}), and slitless spectra in [\ion{O}{3}] $\lambda$5007 
({\it center}) and the triplet of [\ion{N}{2}] $\lambda$6548, H$\alpha$, 
and [\ion{N}{2}] $\lambda$6583 ({\it right}) of the PN targets in the STIS 
survey. The images are 3\arcsec\ on a side with a log intensity stretch 
unless otherwise indicated.  
The LMC nebulae are, from top to bottom, 
J~5, J~33, MG~14 (square-root intensity scale), MG~16, and MG~29. 
The orientation for each image is indicated on the figure, with north 
lying in the direction of the arrow and east to the left. 
\label{Neb_1}}

\figcaption[f5.eps]{
Same as Fig.~\ref{Neb_1} for the LMC nebulae 
MG~4, and Mo~21, except that the images are 
4\arcsec\ $\times$ 4\arcsec. 
\label{Neb_2}}

\figcaption[f6.eps]{
Same as Fig.~\ref{Neb_1} for the LMC nebulae: 
MG~40, MG~45, MG~51, MG~70, and Mo~7. 
\label{Neb_3}}

\figcaption[f7.eps]{
Same as Fig.~\ref{Neb_1} for the LMC nebulae 
Mo~33 (6\arcsec\ $\times$ 6\arcsec), and Mo~47 (4\arcsec\ $\times$ 4\arcsec). 
\label{Neb_4}}

\figcaption[f8.eps]{
Same as Fig.~\ref{Neb_1} for the LMC nebulae 
Mo~36 (square-root intensity scale), Sa~104a, Sa~107, Sa~117, and Sa~121. 
\label{Neb_5}}

\figcaption[f9.eps]{
Same as Fig.~\ref{Neb_1} for the LMC nebulae 
SMP~3, SMP~5, SMP~6 (square-root intensity scale), SMP~11, and SMP~29. 
\label{Neb_6}}

\figcaption[f10.eps]{
Same as Fig.~\ref{Neb_1} for the LMC nebulae 
SMP~14, and SMP~45, except that the images are 
4\arcsec\ $\times$ 4\arcsec. 
\label{Neb_7}}

\figcaption[f11.eps]{
Same as Fig.~\ref{Neb_1} for the LMC nebulae 
SMP~37, SMP~39, SMP~43, SMP~47, and SMP~48. 
\label{Neb_8}}

\figcaption[f12.eps]{
Same as Fig.~\ref{Neb_1} for the LMC nebulae 
SMP~49, SMP~51, SMP~52, SMP~57, and SMP~61. 
\label{Neb_9}}

\figcaption[f13.eps]{
Same as Fig.~\ref{Neb_1} for the LMC nebulae 
SMP~62, SMP~63, SMP~64, SMP~67, and SMP~68. 
\label{Neb_10}}

\figcaption[f14.eps]{
Same as Fig.~\ref{Neb_1} for the LMC nebulae 
SMP~69, SMP~73, SMP~74, SMP~75, and SMP~82. 
\label{Neb_11}}

\figcaption[f15.eps]{
Same as Fig.~\ref{Neb_1} for the LMC nebulae 
SMP~83 (6\arcsec\ $\times$ 6\arcsec), and SMP~91 (4\arcsec\ $\times$ 4\arcsec). 
\label{Neb_12}}

\figcaption[f16.eps]{
Same as Fig.~\ref{Neb_1} for the LMC nebulae 
SMP~84, SMP~88, SMP~89, SMP~92, and SMP~98. 
\label{Neb_13}}

\figcaption[f17.eps]{
Same as Fig.~\ref{Neb_1} for the nebula 
LMC--SMP~101. 
\label{Neb_14}}

\figcaption[f18.eps]{
Same as Fig.~\ref{Neb_1} for the SMC nebulae 
SMP~02, SMP~03, SMP~15, SMP~16, and SMP~28. 
\label{Neb_15}}

\figcaption[f19.eps]{
PNe observed in other {\it HST} programs, on the same spatial scale as for 
Fig.~\ref{Neb_1}, with a square-root intensity stretch. 
The nebulae and spatial extent of the images are: 
LMC--MA~17 (4\arcsec $\times$ 4\arcsec\ with log intensity stretch), 
LMC--MG~60 (4\arcsec\ $\times$ 4\arcsec), 
LMC--Mo~17 (7\arcsec\ $\times$ 6\arcsec), 
LMC--RP~265 (4\arcsec\ $\times$ 4\arcsec), and 
SMC--JD~12 (3\arcsec\ $\times$ 3\arcsec). 
The orientations are indicated next to each image, with north lying in 
the direction of the arrow and east to the left.
\label{Neb_16}}

\end{document}